\documentclass[prb,preprint,12pt,endfloats*]{revtex4}

\usepackage[latin1]{inputenc}
\usepackage[T1]{fontenc}
\usepackage[english]{babel}
\usepackage{ae}
\usepackage{color}
\usepackage{amsmath}
\usepackage{dcolumn}
\usepackage{amssymb}
\usepackage{amsfonts}
\usepackage{braket}
\usepackage{graphicx}
\usepackage{graphics}
\usepackage{lscape}

\begin{document}


\title{Vibrational analysis of methyl cation -- rare gas atom complexes:\\ CH$_3^+$--Rg (Rg=He, Ne, Ar, Kr)}

\author{Jan Meisner, Philipp P.\ Hallmen, Johannes K\"astner and Guntram Rauhut\footnote{To whom correspondence should be addressed:
rauhut@theochem.uni-stuttgart.de}}

\affiliation{Institute for Theoretical Chemistry, University of Stuttgart, Pfaffenwaldring 55, D-70569 Stuttgart, Germany}

\begin{abstract}
The vibrational spectra of simple CH$_3^+$--Rg (Rg=He, Ne, Ar, Kr) complexes have been studied by
vibrational configuration interaction (VCI) theory relying on multidimensional potential energy
surfaces (PES) obtained from explicitly correlated coupled cluster calculations, CCSD(T)-F12a. In agreement with
experimental results, the series of rare gas atoms leads to rather unsystematic results and indicates
huge zero point vibrational energy effects for the helium complex. In order to study these 
sensitive complexes more consistently, we also introduce configuration averaged vibrational self-consistent
field theory (CAVSCF), which is a generalization of standard VSCF theory to several configurations.
The vibrational spectra of the complexes are
compared to that of the methyl cation, for which corrections due to scalar-relativistic effects,
high-level coupled-cluster terms, i.e.\ CCSDTQ, and core-valence correlation have explicitly been
accounted for. The occurrence of tunneling splittings for the vibrational ground-state of CH$_3^+$--He
has been investigated on the basis of semiclassical instanton theory. These calculations and a direct
comparison of the energy profiles along the intrinsic reaction coordinates (IRC) with that of the
hydronium cation, H$_3$O$^+$, suggest that tunneling effects for vibrationally excited states should be
very small.
\end{abstract}

\maketitle
\clearpage

\section{Introduction}
The spectroscopic investigation of ionic complexes is a challenging task and is usually accompanied by 
{\it ab initio} calculations to guide the interpretation of the observed spectra.
\cite{hech3,nech3,arch3,luckhaus,dopfer_review,asvany1,hiraoka} However, 
as these clusters are often very floppy, they request a proper account of anharmonic corrections and correlation effects 
in order to yield reliable results. Complexes of the methyl cation with rare gas atoms
have extensively been studied by Dopfer and co-workers and a wealth of information is available for these 
systems.\cite{hech3,nech3,arch3,luckhaus,dopfer_review} 
These authors used infrared photodissociation spectroscopy in combination with {\it ab initio} 
calculations at the MP2 level with a modified aug-cc-pVTZ basis set to gain an understanding of the 
structural and vibrational properties of these complexes. Of course, 2nd order M{\o}ller--Plesset perturbation 
theory is limited in accuracy and predictions are often of qualitative rather than quantitative nature.
Moreover, the harmonic approximation leaves many questions open and thus an anharmonic treatment of
these systems is highly desirable.
Consequently, Dopfer and co-workers concluded that {\it future efforts to explore the properties of 
the intermolecular interaction in CH$_3^+$--He in a more quantitative way will require a potential 
energy surface that takes monomer relaxation into account}.
This prompted Asvany and collaborators to perform calculations employing 2nd order vibrational perturbation theory, VPT2,\cite{vpt2}
to account for anharmonicity effects in the CH$_3^+$--He complex. These authors used more reliable coupled cluster 
calculations, i.e.\ CCSD(T)/cc-pVTZ, to determine a quartic force field of the potential energy surface.\cite{asvany1} 
The same way, all calculations presented here use high-level explicitly correlated coupled-cluster theory to determine a 
multi-dimensional potential energy surface without imposing any constraints on the coordinates.\\

It was found, that the structures of CH$_3^+$--Rg clusters are strongly dependent on the size of the rare gas atom. For example,
while complexation with helium leads to a very moderate distortion of the CH$_3^+$ moiety, the interaction
with an argon or krypton atom results in a pronounced pyramidalization of the methyl cation. 
{\color{black}For a detailed discussion of that issue see Ref.\ \onlinecite{dopfer_review} and references 
therein.} Besides that, the zero
point vibrational energy leads to large corrections in the He cluster, while the effects are less dramatical for 
Ar or Kr complexes. A proper quantization of these effects requests high-level electronic structure
calculations in combination with high-dimensional variational calculations for determining vibrational
wavefunctions.\\

A particular aspect concerns tunneling splitting within these complexes.\cite{hech3,nech3,arch3,luckhaus,dopfer_review} 
As the dominant interaction
within these complexes is associated with the 2p$_z$ orbital of the planar (D$_{3h}$) methyl cation,
the rare gas atom can be bound to either side of it. {\color{black}This results in two equivalent minima, which are
mainly connected by a movement of the hydrogen atoms - rather than a large amplitude motion of the rare gas 
atom around the CH$_3^+$ moiety.} The formal transition state associated with the minimum energy path is planar and shows C$_{2v}$ 
symmetry. Consequently, tunneling splitting may formally occur in all of these complexes. Based on calculations
on a 3-dimensional energy surface of the interaction potential at the MP2 level (keeping
all intramolecular coordinates of the CH$_3^+$ moiety frozen), significant splittings were 
found for the bending and stretching modes in the CH$_3^+$--He complex and a splitting of about 0.008 cm$^{-1}$ 
for the vibrational ground state.\cite{luckhaus} In contrast to that recent studies by T\"opfer {\it et al.} based 
on double resonance rotational spectroscopy do not confirm this result.\cite{asvany2} Therefore, we will consider 
this aspect in some detail. \\

Within the course of this study, we introduce and use configuration-averaged vibrational self-consistent field
theory (CAVSCF) to determine optimized one-mode wavefunctions in order to avoid symmetry-broken solutions
in the subsequent vibrational configuration interaction calculations (VCI). 
This concept has been developed many years ago
within electronic structure theory and has been revived within the context of calculating
crystal field splittings for molecular magnets.\cite{zerner,soncini,phil1} However, this theory has not yet been transferred to 
the solution of the vibrational Schr\"odinger equation. For that reason, we will outline the basics of this theory 
briefly.

\section{Configuration averaged VSCF theory}
In order to keep the equations as simple as possible, we will neglect the vibrational angular
momentum terms\cite{convergence} and the Watson correction term as occurring in the Eckart--Watson Hamiltonian.\cite{watson}
In all what follows the multidimensional potential energy surface shall be represented by an
$n$-mode representation\cite{vreview} expressed in polynomials, i.e.\
\begin{equation}
V{({\bf q})} = V_{1D} + V_{2D} + V_{3D} + \dots 
\end{equation}
with
\begin{eqnarray}
V_{1D} & = & \sum_i \sum_r p^{(i)}_{r} q_i^r \\
V_{2D} & = & \sum_{i<j} \sum_{rs} p^{(ij)}_{rs} q_i^rq_j^s \\
V_{3D} & = & \dots 
\end{eqnarray}
$q_i^r$ denotes a polynomial of order $r$ for mode $i$, while $p$ are the associated coefficients
as obtained from fitting procedures.\cite{fit} Note, by definition $p^{(ii)}_{rs}=0$, $p^{(iij)}_{rst}=0$ etc.
Within VSCF theory the wave function is given as a 
product of modals $\varphi_i^I$ belonging to configuration $I$:
\begin{equation}
\Psi^I = \prod_i \varphi_i^I(q_i)  \qquad \mbox{with} \qquad \varphi_i^I = \sum_\mu C_{\mu i}^I \chi_{\mu i}
\end{equation}
$\chi_{\mu i}$ are the basis functions, e.g.\ harmonic oscillator function or distributed Gaussians, and
$C_{\mu i}^I$ are the modal coefficients to be optimized within the VSCF iterations. We will employ the
integrals
\begin{equation}
Q_{\mu\nu}^{ir} = \left < \chi_{\mu i} \left | q_i^r \right | \chi_{\nu i} \right > 
\quad \mbox{and} \quad X_{ir}^I = \sum_{\mu\nu} C_{\mu i}^IC_{\nu i}^I Q_{\mu\nu}^{ir}
\end{equation}

\subsection{Standard VSCF Theory}
Using the $n$-mode representation of the potential and the aforementioned definitions, the energy
expectation value within VSCF theory\cite{bowman1,bowman2,gerber1} can be written as
\begin{eqnarray}
E_I & = & \sum_i\sum_{\mu\nu}C_{\mu i}^IC_{\nu i}^I \left [ 
-\frac{1}{2}T_{\mu\nu}^i + \sum_r Q_{\mu\nu}^{ir} \left [ p_r^{(i)} + \right . \right . \nonumber \\
&& \left . \left . \sum_{j} \sum_s X_{js}^I \left [ \frac{1}{2} p_{rs}^{(ij)} + \dots \right ] \right ] \right ]
\label{ei}
\end{eqnarray}
Herein $T_{\mu\nu}^i$ denotes the kinetic energy associated with mode $i$ in the basis of the primitive
functions. As these functions are mode-dependent, the integral of the kinetic energy also depends on the
mode.  Requesting normality of all modals for each mode yields the Lagrangian
\begin{equation}
L = E_I - \sum_i \varepsilon_i^I \left (\left < \varphi_i^I | \varphi_i^I \right > - 1 \right ) 
\end{equation}
Variation of the Lagrangian with respect to the modal coefficients finally leads to a generalized
eigenvalue equation with the hermitian Fock-type matrix
\begin{equation}
F^{i,I}_{\mu\nu} = - \frac{1}{2}T_{\mu\nu}^i + \sum_r Q_{\mu\nu}^{ir} \bar{p}_r^{(i),I}
\end{equation}
where $\bar{p}_r^{(i),I}$ denotes the effective potential given as
\begin{equation}
\bar{p}^{(i),I}_r = p_r^{(i)} + \sum_j \sum_s X_{js}^I \left [ p_{rs}^{(ij)} + \sum_k\sum_t \left [
\frac{1}{2} p_{rst}^{(ijk)} + \dots \right ] \right ] 
\end{equation}
As the Fock-type matrix ${\bf F}$ depends on the modal coefficients, the equations must be solved
iteratively until self-consistency is achieved, which is responsible for the name of this procedure.

\subsection{CAVSCF Theory}
In order to optimize the modals for degenerate states in molecules belong to non-Abelian point groups,
VSCF theory can be generalized to several configurations by simple averaging.
Assuming identical weights for all configurations to be considered in CAVSCF, the energy expression is
given as
\begin{equation}
E_\text{CA}=\frac{1}{N}\sum_I E_I
\end{equation}
where $N$ is the number of configurations to be considered. We further introduce a fractional
occupation number $\gamma_{a_i}$ for modal $a_i$ of mode $i$
\begin{equation}
\gamma_{a_i} = \frac{n_{a_i}}{N}
\end{equation}
where $n_{a_i}$ is the number of configurations, in which the modal $a_i$ of mode $i$
is used. The condition
\begin{equation}
\sum_{a_i}\gamma_{a_i} = 1
\end{equation}
must hold for every mode $i$. Clearly, the weight factors $\gamma$ simply depend on the
configurations to be considered simultaneously. With this, eq.\ \ref{ei} can be generalized to
\begin{eqnarray}
E_\text{CA}&=&\sum_i\sum_{a_i}\gamma_{a_i}\sum_{\mu\nu} C_{\mu i}^{a_i}C_{\nu i}^{a_i} \left [ - 
\frac{1}{2} T_{\mu\nu}^i + \sum_r Q_{\mu\nu}^{ir} \left [ p_r^{(i)} + \right . \right . \nonumber \\
&& \left . \left . \sum_j \sum_{b_j} \gamma_{b_j} 
\sum_s X_{js}^{b_j} \left [ \frac{1}{2} p_{rs}^{(ij)} + \dots \right ] \right ] \right ]
\end{eqnarray}
Likewise, the Lagrangian can now be written by extending it to all active modals
within each mode $i$:
\begin{equation}
L_\text{CA}=E_\text{CA}-\sum_i \sum_{a_ib_i} \varepsilon_i^{a_ib_i} \left ( \left < \varphi_i^{a_i} | \varphi_i^{b_i}
\right > - \delta^i_{a_ib_i} \right )
\end{equation}
Variation of the Lagrangian finally yields a Fock-type matrix given as
\begin{eqnarray}
F_{\mu\nu}^{i,\text{CA}} & = & -\frac{1}{2} T_{\mu\nu}^i + \sum_r Q_{\mu\nu}^{ir} \left [ p_r^{(i)} + \right . \\
&& \left . \sum_j \sum_{b_j} \gamma_{b_j} \sum_s X_{js}^{b_j} \left [ \frac{1}{2} p_{rs}^{(ij)} 
+ \dots \right ] \right ] \nonumber
\end{eqnarray}
Comparison of the equations for CAVSCF and VSCF theory reveals, that besides the introduction
of fractional occupations $\gamma_{a_i}$ it is simply an additional loop over the
active modals and the extension of the contracted integrals ${\bf X}$ to all active modals rather than just one
modal in standard VSCF theory. However, this does not lead to any storage problems. Consequently,
CAVSCF theory is a rather simple generalization of VSCF theory, which is just the special
case for one configuration. Using identical weights for the individual configurations of degenerate 
states, CAVSCF can be used for determining state-specific solutions without symmetry breaking within
the calculation of degenerate states in non-Abelian molecules.

\section{Instanton Theory}
To quantify the tunneling splitting of the CH$_3^+$--He cluster in the vibrational ground state, 
we used semiclassical instanton theory.\cite{mil75,vai82,mil08a,ric11}
In instanton theory, the tunneling splitting $\Delta$ is calculated by the ratio
of the partition function of the full quantum system, $Q(\beta)$,
and the partition function of the non-tunneling system,
$Q_0(\beta)$
\begin{equation}
  \label{eq:instanton}
  \lim_{\beta \rightarrow \infty} \frac{Q_{\text{}}(\beta)}{Q_0(\beta)}
  = \cosh\left({\frac{\beta \Delta}{2}}\right),
\end{equation}
where 
$\beta = \frac{1}{k_{\text{B}} T}$ is the inverse temperature and $k_{\text{B}}$ is Boltzmann's constant.
To calculate the partition function of the full system, $Q_{\text{}}(\beta)$, in principle all paths from one minimum to the other  
should be considered following Feynman's idea of path integrals.
However, in instanton theory the most likely tunneling path, the so-called instanton, is optimized and 
the steepest-descent integration around it is used.
Thus, quantum fluctuations about this path are considered in harmonic approximation by diagonalization of the Hessian matrix of the full instanton.
Along the tunneling path, full anharmonicity is included, and, thus, 
instanton theory can be considered to 
be correct by an order of magnitude. \cite{richardson2011}

\section{Computational details}
Explicitly correlated coupled-cluster calculations, CCSD(T)-F12a and CCSD(T)-F12b,\cite{ccf12} in combination with an augmented 
triple-$\zeta$ basis, i.e.\ aug-cc-pVTZ,\cite{vtz} have been used to determine the structures of all complexes 
and the bare methyl cation, as well as for the instanton calculations. Note that we used F12a theory for the calculation 
of all clusters, but employed F12b calculations for the monomer for which we also used larger basis sets.
In general F12b theory shows slightly better performance for large basis sets than the F12a approximation.
These levels can roughly be compared with conventional coupled-cluster 
calculations relying on an augmented quintuple-$\zeta$ basis set.\cite{knizia09} At the same level of theory,
harmonic frequencies and normal coordinates as needed for spanning the potential 
energy surface have been determined. For all calculations we used the {\sc Molpro} package of {\it ab initio}
programs.\cite{MOLPRO_dev_brief}\\

We used $n$-mode representations\cite{vreview} of the potential energy surfaces truncated after the 4-mode coupling terms 
within the vibrational configuration interaction calculations.\cite{neff1,neff2} Within all these 
calculations the expansion points of the $n$-mode representations were chosen to be the 
equilibrium structures, i.e.\ local PESs were generated, which harbor just one of the two possible 
minima. {\color{black}A multi-level scheme
has been exploited, in which the 1-mode and 2-mode coupling terms were determined at the CCSD(T)-F12a/aug-cc-pVTZ level, 
while the 3-mode and 4-mode coupling terms were obtained from CCSD(T)-F12a/aug-cc-pVDZ theory. 
In order to exploit symmetry within the PES generation more efficiently, the displacement vectors 
of the degenerate normal coordinates were rotated by $\pi$/4 as described most recently.\cite{benni}} The 
resulting grid representation of the potential was transformed into an analytical representation
based on polynomials up to 7th order. For the transformation we used our recently developed algorithm
relying on Kronecker products.\cite{fit} Initial vibrational wavefunctions were obtained from vibrational
self-consistent field theory employing a mode-specific basis of 20 distributed Gaussians. 
For this, the Watson Hamiltonian\cite{watson} has been used, in which the $\mu$-tensor was truncated after the 
0th order term.\cite{convergence} The subsequent VCI calculations were based on Hartree products (configurations)
generated from the VSCF modals (one-mode wavefunctions) of the vibrational ground-state. The correlation
space within the VCI calculations was limited to 6-tuple
excitations up to the 7th root and a maximum sum of quantum numbers of 25. A  
configuration-selective VCI approach\cite{neff1,neff2} based on VSCF modals of the vibrational ground-state or
modals obtained from CAVSCF theory have been used to limit the computational effort in this step.
Symmetry breakings due to insufficiencies in the potential, fitting to the polynomials or 
limitations of the configuration space etc.\ were found to be as small as 0.01 cm$^{-1}$. \\

Instantons were calculated at 2.0~K and 1.75~K using a quasi-Newton--Raphson algorithm.\cite{rommel2011}
{\color{black} Note that for the tunneling splitting, the limit of zero temperature is required, cf.\ Eq.\ \ref{eq:instanton}.
Here, the temperature is a parameter rather than a physical quantity.}
610 images were used for the discretization of the instanton path.
Instanton paths were considered to be converged when the maximum component of the gradient (in atomic units, i.e. $m_\text{e}=1$) was below $10^{-8}$.
The rotational contribution to the partition function of the instanton and the minimum structure was approximated as classical rigid rotors.
For the calculation of the intrinsic reaction coordinate (IRC) and the the instanton calculations 
the DL-Find optimization library was used.\cite{kaestner2009}
The communication between DL-Find and {\sc Molpro}\cite{MOLPRO_dev_brief} was realized using the Chemshell interface.\cite{metz2014}
For the IRC optimization, the Hessian predictor-corrector algorithm\cite{hratchian2004,hratchian2005} 
has been applied as recently implemented in DL-Find.\cite{meisner2017b}
A step size of 0.02 a.u.\ 
was used to optimize the IRC on the comparably flat PES.
To obtain a smooth gradient, we tightened the SCF and CCSD(T) convergence to $10^{-11}$~Hartree
for the calculation of the IRC, instantons, and Hessians along the instanton pathways.

\section{Results and discussion}
\subsection{Geometrical parameters}
\begin{figure}
\includegraphics[width=0.45\textwidth]{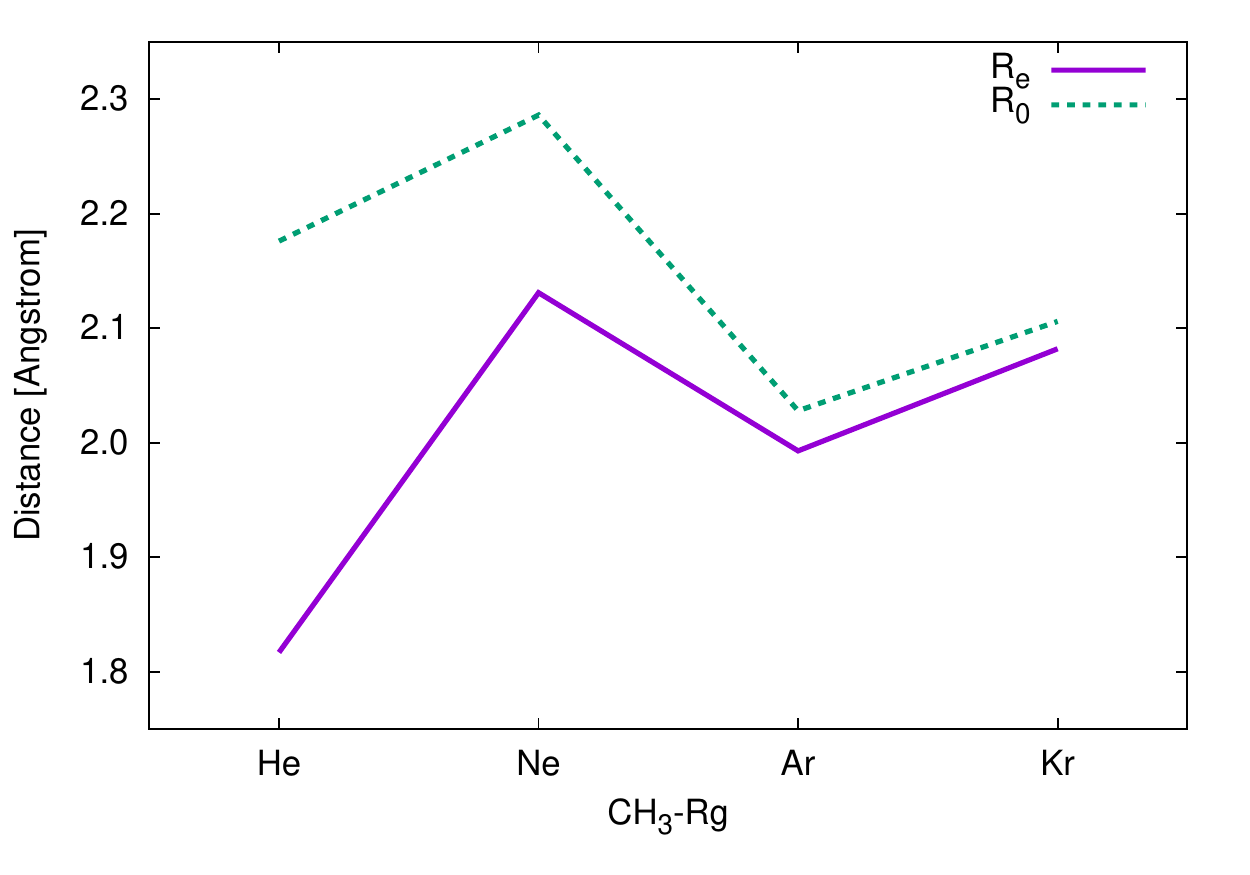}
\caption{Equilibrium (purple, solid) and vibrationally averaged (green, dotted) distances of the rare gas atoms to the carbon atom in 
CH$_3^+$--Rg complexes (Rg=He, Ne, Ar, Kr). \label{dist}}
\end{figure}

While the equilibrium structure of a molecule is uniquely defined, there are several possibilities 
concerning vibrationally averaged structures. Usually, the definitions differ with respect to 
the operator within the integral of the expectation value. We employed internal coordinates, i.e.\ bond
lengths etc., which were expanded in terms of normal coordinates, the latter also being used as variables for the 
wavefunctions. Temperature effects by Boltzmann weighting were not included in these calculations. 
Structural parameters being most important for this study here are listed in Table \ref{structures}.
\begin{table*}
\caption{Structural parameters of CH$_3^+$--Rg (Rg=He, Ne, Ar, Kr) complexes. Values are given in {\AA}ngstr{\o}m or 
degree.\label{structures}}
\begin{center}
\begin{tabular}{lcrrccccccccccc}
\hline \hline \\[-1ex]
Complex && \multicolumn{2}{c}{$\theta_\text{e}$(HCRg)$^a$} && \multicolumn{2}{c}{$R_\text{e}$(C--Rg)$^a$} && 
           \multicolumn{2}{c}{$R_0$(C--Rg)$^a$} && \multicolumn{2}{c}{$\Delta R_\text{e0}$(C--Rg)$^{a,b}$} && $\Delta R_\text{e}$(C--H$_3$)$^c$ \\
\hline \\[-1ex]
CH$_3^+$--He &&  91.51 & (91.4) && 1.817 & (1.834) && 2.178 & (2.176) && 0.361 & (0.34) && 0.029 \\
CH$_3^+$--Ne &&  91.31 & (91.4) && 2.131 & (2.135) && 2.286 & (2.300) && 0.155 & (0.17) && 0.025 \\
CH$_3^+$--Ar &&  98.83 & (99.0) && 1.993 & (1.988) && 2.030 & (2.053) && 0.038 & (0.10) && 0.166 \\
CH$_3^+$--Kr && 100.51 &        &&  2.082 &        && 2.106 &         && 0.025 &        && 0.198 \\
\hline \hline
\end{tabular} 
\end{center}
\vspace*{1mm}
{$^a$ \footnotesize Values in parentheses were taken from Ref.\ \onlinecite{dopfer_review} and refer to
MP2/aug-cc-pVTZ calculations. Note that our $R_0$(C--Rg) and $\Delta R_\text{e0}$(C--Rg) values 
rely on a different definition than Dopfer's $R_{cm}$ and $R_{cm}-R_e$ values given in parentheses. $\hfill$}\\
{$^b$ \footnotesize Difference of the C--Rg bond length between the equilibrium 
and the vibrationally averaged structure. $\hfill$}\\ 
{$^c$ \footnotesize Out-of-plane distortion of the carbon atom relative 
to the plane defined by the three hydrogen atoms within the equilibrium structure. $\hfill$} \\
\end{table*}
According to Table \ref{structures} and
as discussed in detail by Dopfer and co-workers\cite{arch3} the interaction of the Ar and Kr atoms with the methyl
cation are much stronger than for the other two systems, 
which leads to a pronounced pyramidalization of the CH$_3^+$ moiety in contrast
to the helium or neon complexes. However, this effect appears to be less pronounced by the coupled-cluster
calculations presented here in comparison to the MP2 results of the literature.\cite{hech3,nech3,arch3,dopfer_review} Our CCSD(T)-F12a results 
also indicate a weaker pyramidalization due to neon than to helium, which is not supported by the 
MP2 results. However, the effects are rather weak and surely within the error bar of the MP2 calculations.
Nevertheless, this result is counter-intuitive as neon shows a larger polarizability than helium and 
thus one would expect a stronger pyramidalization. The calculated interaction energy of the neon-complex is also computed to 
be larger than for helium (see below). However, due to the bigger covalent radius, the C--Ne distance is significantly
longer than the C--He distance in the helium complex and thus there appears to be a sensitive equilibrium between the 
contributing physical forces. The differences between the vibrationally averaged C--Rg distances, $R_0$(C-Rg),
and the corresponding values within the equilibrium structures, i.e.\ $\Delta R_\text{e0}$(C--Rg), 
are in nice agreement with the estimates of Dopfer and co-workers (cf.\ the comparison in Table \ref{structures}).\cite{hech3,nech3} 
Note that these values cannot directly
be compared as Dopfer and co-workers determined the difference in a pseudo-diatomic system and thus these
authors refer to the center of mass of the CH$_3^+$ moiety, while our values refer to the true atom-atom
distances. As the pyramidalization is largest for the Ar and Kr clusters, the largest difference between our 
$\Delta R_\text{e0}$(C--Rg) values 
and the related results of Dopfer and co-workers must arise for these particular systems. Table \ref{structures} clearly
supports the known result, that huge ZPVE effects can be found for the CH$_3^+$--He complex. The unsystematic
behavior of the C--Rg distances is visualized in Figure \ref{dist}.
As must be expected, the ZPVE effects are less pronounced in the deuterated CD$_3^+$--Rg complexes. The 
corresponding values are listed in Table \ref{deutstruct}.
\begin{table}
\caption{Structural parameters of CD$_3^+$--Rg (Rg=He, Ne, Ar, Kr) complexes. Values are given in {\AA}ngstr{\o}m.
\label{deutstruct}}
\begin{center}
\begin{tabular}{lcrcccc}
\hline \hline \\[-1ex]
Complex && $R_0$(C--Rg) & $\Delta R_\text{e0}$(C--Rg)$^a$  \\
\hline \\[-1ex]
CD$_3^+$--He && 2.102 & 0.285 \\
CD$_3^+$--Ne && 2.257 & 0.122 \\
CD$_3^+$--Ar && 2.023 & 0.030 \\
CD$_3^+$--Kr && 2.102 & 0.020 \\
\hline \hline
\end{tabular} 
\end{center}
\vspace*{1mm}
{\raggedright $^a$ \footnotesize Difference of the C--Rg bond length between the equilibrium
and the vibrationally averaged structure.}\\
\end{table}

\subsection{Energetic considerations}
Complexation energies, $E_\text{Cplx}$, relaxation energies, $E_\text{Relax}$, and dissociation energies,
$D_\text{e}=E_\text{Cplx}-E_\text{Relax}$, of the CH$_3^+$--Rg 
complexes are shown in Table \ref{energies}. 
\begin{table*}
\caption{Interaction energies of CH$_3^+$--Rg (Rg=He, Ne, Ar, Kr). Values are given in cm$^{-1}$.\label{energies}}
\begin{center}
\begin{tabular}{lcrrrrrrrrrr}
\hline \hline  \\[-1ex]
Complex      && \multicolumn{2}{c}{$E_\text{Cplx}^a$} && \multicolumn{2}{c}{$E_\text{Relax}^a$} 
             && \multicolumn{2}{c}{$D_\text{e}^a$} && $D_0$\\
\hline \\[-1ex]
CH$_3^+$--He &&   886.7 &  (761.6) &&   61.1 &   (54.5) &&  825.6 &  (707.1) &&  247.1 \\
CH$_3^+$--Ne &&  1241.5 & (1011.2) &&   46.8 &   (57.7) && 1194.7 &  (958.5) &&  663.7 \\
CH$_3^+$--Ar &&  8650.7 & (7781.6) && 2073.0 & (2206.0) && 6577.7 & (5575.6) && 5428.4 \\
CH$_3^+$--Kr && 12579.6 &          && 2929.9 &          && 9649.7 &          && 8486.2 \\
\hline \hline
\end{tabular}
\end{center}
\vspace*{1mm}
{$^a$ \footnotesize Values in parentheses were taken from Ref.\ \onlinecite{dopfer_review} and refer to
MP2/aug-cc-pVTZ calculations.}
\end{table*}
Clearly, the complexation energies and the dissociation energies grow with increasing period of the rare gas atoms.
However, this does not hold true for the relaxation energy of the CH$_3^+$ moiety, which is in agreement with the 
pyramidalization as discussed above, i.e.\ for the neon complex the pyramidalization (see $\Delta R_\text{e}$(C--H$_3$) in Table \ref{structures})
is less pronounced than for helium and likewise is the relaxation energy. Moreover, as discussed in detail by Dopfer and 
co-workers, the ZPVE has a huge effect on the dissociation energy of CH$_3^+$--He, which steadily decreases with the period
of the rare gas atoms. All energies listed in Table \ref{energies} are much larger for Ar and Kr than for He and Ne. This is
in line with the conclusion of Dopfer and co-workers that {\it only the CH$_3^+$--Rg dimers with large rare gas ligands 
have bonds with substantial covalent character}. {\color{black}This result is also supported by natural population and bond 
analyses (NBO),\cite{nbo} which show a covalent C-Rg bond for Rg=$\{$Ar,Kr$\}$ associated with a substantial charge transfer.
For the small rare gas atoms, i.e.\ He and Ne, the NBO and the charge transfer cannot be seen. An analysis based on
symmetry adapted perturbation theory (SAPT)\cite{sapt} reveals, that in the latter two systems, the interaction energy is
dominated by induction and to a lesser extent by dispersion. Again, this agrees with the conclusions of Dopfer.\cite{dopfer_review}
} For the deuterated species, the $D_0$ values are 348.8 (He), 785.2 (Ne), 5717.3 (Ar) and 
8781.4 cm$^{-1}$ (Kr).  \\

As mentioned above, the complexes discussed here allow for an internal rotation of the CH$_3^+$ moiety, which
give rise to formal planar transition states of C$_{2v}$ symmetry. We have computed these barriers to be 
698.5 (He), 916.4 (Ne), 5891.7 (Ar) and 8763.8 cm$^{-1}$ (Kr). {\color{black}Dopfer and co-workers 
reported these values to be about 600, 750, and 6000 cm$^{-1}$ (He, Ne, Ar).\cite{nech3}}
Clearly, the barriers for CH$_3^+$--Ar and 
CH$_3^+$--Kr are too high to show any impact on the vibrational structure. It is the covalent bond character,
which hinders the internal rotation in the latter two complexes. However, for the helium and neon complexes,
the barriers are just slightly below the dissociation channels ($D_\text{e}$) and are low enough
that tunneling effects may lead to vibrational splittings as discussed by Dopfer and 
Luckhaus.\cite{luckhaus} We will discuss this aspect in detail below.


%

\subsection{Vibrational frequencies}
\subsubsection{Methyl cation}
The experimental determination of the fundamental modes of the bare methyl cation (D$_{3h}$) is a non-trivial task
and several attempts have been undertaken to assign the individual modes.\cite{crofton,koenig1,koenig2,dyke,liu}
The only accurately determined frequency of the methyl cation is the degenerate CH stretching mode, which has been determined 
by Crofton {\it et al.}\cite{crofton} using infrared spectroscopy to be 3108.4 cm$^{-1}$. The only other mode, which
has been determined experimentally, is the out-of-plane umbrella mode $\nu_2$. From photoelectron spectroscopy
this vibration was found at 1406$\pm$30, 1366$\pm$20 and 1356 cm$^{-1}$, respectively, 
while ion pair imaging spectroscopy yielded a value of 1359$\pm$7 cm$^{-1}$.\cite{dyke,koenig1,koenig2,liu}
A more recent measurement by Cunha de Miranda {\it et al.}\cite{pbotsch} based on
threshold photoelectron spectroscopy determined this fundamental at 1387$\pm$15 cm$^{-1}$. The scattering 
of these values and the large error bars indicate the problems associated with the assignment of this particular 
band. For CD$_3^+$ the situation differs completely. Asvany {\it et al.}\cite{asvany1} determined all fundamentals except the 
symmetric CD stretching mode, which are in nice agreement with our theoretical predictions (see below).\\
 
Due to its limited size, the methyl cation has theoretically been studied by many 
authors.\cite{sears,keceli,pbotsch,ch3harm} 
Yu and Sears computed the anharmonic vibrational frequencies of the methyl cation and its 
deuterated isotopologue using CCSD(T) calculations at the basis set limit.\cite{sears} 
These authors applied a coordinate scaling procedure in order to compensate for systematic overestimations 
in their calculations, which was criticized in the work of Cunha de Miranda {\it et al.}\cite{pbotsch}
Moreover, Yu and Sears used a rather limited force field in contrast to the calculations of Keceli {\it et al.},\cite{keceli}
Cunha de Miranda {\it et al.}\cite{pbotsch} and our own calculations, which include some 11248 potential terms. 
The results of Keceli {\it et al.}\cite{keceli} do not include corrections due to the vibrational angular momentum (VAM)
terms, which were found to be very important by Cunha de Miranda {\it et al.},\cite{pbotsch} who determined corrections 
by up to 22 cm$^{-1}$. Therefore, we consider the most reliable values to be those of Cunha de Miranda {\it et al.},\cite{pbotsch}
which essentially coincide with our own calculation. However, this must be expected as the setup
of our calculations essentially is the same as theirs although we use an $n$-mode expansion truncated 
after the 4-mode coupling terms rather than the 3-mode couplings. However, this simply indicates that
the $n$-mode expansion of the potential energy surface is converged after the inclusion of the 3-mode 
coupling terms. Our results are summarized in Table \ref{ch3freq}.
\begin{table*}
\caption{Harmonic and anharmonic vibrational frequencies of CH$_3^+$ and CD$_3^+$.
Values are given in cm$^{-1}$.\label{ch3freq}}
\begin{center}
\begin{tabular}{clcrrrrcrrr}
\hline \hline \\[-1ex]
&&& \multicolumn{4}{c}{CH$_3^+$} && \multicolumn{3}{c}{CD$_3^+$} \\
\cline{4-7} \cline{9-11} \\[-1ex]
Mode & Sym. && Harm.\ & VCI$^a$ & VCI$^b$ & Exp.$^c$ && Harm.\ & VCI$^a$ & Exp.$^d$ \\
\hline \\[-1ex]
ZPVE    & $a'_1$     && 6908.6 & 6807.5 & 6816.6 &        && 5101.3 & 5046.0 &      \\
$\nu_1$ & $a'_1$     && 3043.4 & 2940.2 & 2943.7 &        && 2153.1 & 2095.7 &      \\
$\nu_2$ & $a''_2$    && 1424.5 & 1405.0 & 1406.7 &        && 1103.6 & 1091.3 & 1090 \\
$\nu_3$ & $e'$       && 3244.0 & 3104.9 & 3109.8 & 3108.4 && 2422.2 & 2343.1 & 2337 \\
$\nu_4$ & $e'$       && 1430.7 & 1393.6 & 1395.9 &        && 1050.7 & 1030.0 & 1027 \\
\hline \hline
\end{tabular}
\end{center}
\vspace*{1mm} {
$^a$ \footnotesize Obtained from CCSD(T)-F12b/cc-pVTZ-f12 calculations.$\hfill$ \\
$^b$ \footnotesize Obtained from CCSD(T)-F12b/cc-pVQZ-f12 calculations and additional
corrections for core-valence electron correlation, 
scalar-relativistic effects and high-order coupled-cluster terms (see text).$\hfill$  \\
$^c$ \footnotesize Experimental data taken from Ref.\ \onlinecite{crofton}$\hfill$ \\
$^d$ \footnotesize Experimental data taken from Ref.\ \onlinecite{asvany1}$\hfill$}
\end{table*}

In order to investigate the effect of core-valence correlation
etc.\ on this small molecule, which has not yet been done in any of the preceding theoretical studies,
we have performed additional calculations, which account for these contributions.
In order to limit the computational effort within the generation of the potential energy surfaces, we
used a multi-level scheme, i.e.\ these further corrections were applied to the 1D and 2D-terms of the $n$-mode 
expansion of the potential, while the 3D and 4D-terms were determined at the CCSD(T)-F12b/cc-pVTZ-f12 level
only.  Core-valence corrections were determined from conventional CCSD(T)/cc-pCVQZ calculation relative to
frozen core CCSD(T)/cc-pVQZ results. This correction was found to be largest for the two CH-stretching modes $\nu_1$ and 
$\nu_3$ and accounted for 4.8 and 4.9 cm$^{-1}$, respectively. For the other two modes the correction was less
than 2.0 cm$^{-1}$. Relativistic effects were determined from scalar-relativistic calculations employing
a 2nd order Douglas--Kroll--Hess Hamiltonian.\cite{dkh1,dkh2} In these calculations a aug-cc-pwCVQZ-dk basis set has been
used. These corrections were determined to be very small, they amounted to less than 0.6 cm$^{-1}$ for all
frequencies. In a last step we incorporated high-order coupled-cluster terms
by using the {\sc GeCCo} suite
of {\it ab initio} programs.\cite{gecco1,gecco2} As these calculations are very demanding, we used a  
standard cc-pVTZ basis set for CCSDT calculations and a cc-pVDZ basis within the CCSDTQ calculations. 
The calculations led to a lowering of the frequencies by up to $-$2.2 cm$^{-1}$. Besides these 
corrections to the potential energy surface, we also included linear terms in the expansion of 
the $\mu$-tensor rather than the constant term only within the evaluation of the VAM corrections.\cite{convergence}
However, these correction were found to be very small and accounted to not more than 0.4 cm$^{-1}$. 
Adding all these 
corrections yields our most reliable values for CH$_3^+$, which are shown in the 3rd column of Table \ref{ch3freq}.
Apparently, the inclusion of core-valence correlation provides the largest correction, which is partly 
compensated for by the high-level coupled-cluster terms. 
This effect is well-known and has been observed by several authors.\cite{highorder1,highorder2,meier}
The comparison of these results with our standard 
calculations at the CCSD(T)-F12b/cc-pVTZ-f12 level shows that the differences are fairly small,
but relies on the error compensation of the aforementioned corrections. \\

For the deuterated isotopologue our VCI calculations are in excellent agreement with the experimental results of 
Asvany {\it et al.}\cite{asvany1} and the theoretical predictions of Cunha de Miranda {\it et al.}\cite{pbotsch}

\subsubsection{CH$_3^+$--Rg complexes}
Harmonic and anharmonic vibrational frequencies for all complexes
are summarized in Tables \ref{hfreq} and \ref{dfreq}.
\begin{table*}
\caption{Harmonic and anharmonic vibrational frequencies of CH$_3^+$--Rg (Rg=He, Ne, Ar, Kr). Value are given in cm$^{-1}$.\label{hfreq}} 
\rotatebox{90}{
\begin{tabular}{clcrrrrrcrrrcrrrcrr}
\hline \hline \\[-1ex]
&&& \multicolumn{5}{c}{CH$_3^+$--He} && \multicolumn{3}{c}{CH$_3^+$--Ne} && \multicolumn{3}{c}{CH$_3^+$--Ar} && \multicolumn{2}{c}{CH$_3^+$--Kr}\\
\cline{4-8} \cline{10-12} \cline{14-16} \cline{18-19} \\[-1ex]
Mode & Sym. && Harm.\ & 3D$^a$ & VPT2$^b$ & VCI & Exp.$^c$ && Harm. & VCI & Exp.$^d$ && Harm. & VCI & Exp.$^e$ && Harm. & VCI \\
\hline \\[-1ex]
ZPVE    & $a_1$     &&  7649.4 &        &      & 7386.0 &        && 7508.9 & 7338.5 &        && 8081.6 & 7956.8 &          && 8088.7 & 7971.0 \\
$\nu_1$ & $a_1$     &&  3058.6 &        & 2935 & 2944.5 & 2946.4 && 3060.1 & 2949.9 &        && 3091.0 & 2975.4 & 2979$^f$ && 3093.2 & 2979.6 \\
$\nu_2$ & $a_1$     &&  1408.2 &        & 1391 & 1383.8 & 1402   && 1406.8 & 1380.4 &        && 1360.2 & 1329.4 &          && 1336.9 & 1304.6 \\
$\nu_s$ & $a_1$     &&   216.6 &   83   &  122 &  138.3 &        &&  182.0 &  147.0 &        &&  374.4 &  338.7 &          &&  402.1 &  377.2 \\ 
$\nu_3$ & $e$       &&  3260.9 &        & 3104 & 3113.5 & 3115.0 && 3262.1 & 3119.3 & 3119.4 && 3271.0 & 3129.9 & 3145(30) && 3264.1 & 3123.2 \\
$\nu_4$ & $e$       &&  1429.3 &        & 1395 & 1393.4 & 1383   && 1431.4 & 1393.5 &        && 1437.9 & 1397.0 &          && 1442.5 & 1402.9 \\
$\nu_b$ & $e$       &&   617.5 & 128/149 &  237 & (322)$^g$&      &&  491.0 & (375)$^g$ &     &&  959.9 &  917.0 &          &&  966.1 &  939.4 \\ 
\hline \hline 
\end{tabular}
} \\
\vspace*{1mm}{
$^a$ \footnotesize Variational 3D frequencies obtained from MP2/aug-cc-pVTZ calculations. Taken from Ref.\ \onlinecite{luckhaus}. $\hfill$\\
$^b$ \footnotesize Anharmonic VPT2 frequencies based on CCSD(T)/aug-cc-pVTZ calculations, taken from Ref.\ \onlinecite{asvany1}. $\hfill$ \\
$^c$ \footnotesize Experimental data taken from Refs.\ \onlinecite{asvany1,asvany2}. $\hfill$\\ 
$^d$ \footnotesize Experimental data taken from Ref.\ \onlinecite{nech3}. $\hfill$\\ 
$^e$ \footnotesize Experimental data taken from Ref.\ \onlinecite{arch3}. $\hfill$\\ 
$^f$ \footnotesize Estimated from a local mode-coupled Morse oscillator model. $\hfill$\\ 
$^g$ Numerically not stable, see text.$\hfill$}
\end{table*}
\begin{table*}
\caption{Harmonic and anharmonic vibrational frequencies of CD$_3^+$--Rg (Rg=He, Ne, Ar, Kr). Value are given in cm$^{-1}$.\label{dfreq}}
\begin{center}
\begin{tabular}{clcrrcrrcrrcrr}
\hline \hline \\[-1ex]
&&& \multicolumn{2}{c}{CD$_3^+$--He} && \multicolumn{2}{c}{CD$_3^+$--Ne} && \multicolumn{2}{c}{CD$_3^+$--Ar} && \multicolumn{2}{c}{CD$_3^+$--Kr}\\
\cline{4-5} \cline{7-8} \cline{10-11} \cline{13-14} \\[-1ex]
Mode & Sym. && Harm.\ & VCI && Harm. & VCI && Harm. & VCI && Harm. & VCI \\
\hline \\[-1ex]
ZPVE    & $a_1$     &&  5677.6 & 5522.8 && 5552.4 & 5455.5 && 5975.8 & 5906.4 && 5979.1 & 5914.3 \\
$\nu_1$ & $a_1$     &&  2164.4 & 2092.4 && 2164.9 & 2092.2 && 2192.9 & 2144.9 && 2196.8 & 2147.2 \\
$\nu_2$ & $a_1$     &&  1078.2 & 1073.3 && 1079.8 & 1031.8 && 1011.3 &  995.5 &&  993.6 &  975.7 \\ 
$\nu_s$ & $a_1$     &&   213.8 &  142.0 &&  174.8 &  141.1 &&  364.8 &  336.9 &&  386.9 &  367.9 \\ 
$\nu_3$ & $e$       &&  2435.7 & 2350.6 && 2436.4 & 2353.9 && 2440.1 & 2359.1 && 2433.1 & 2353.4 \\
$\nu_4$ & $e$       &&  1050.1 & 1030.3 && 1052.5 & 1031.8 && 1052.5 & 1032.7 && 1053.6 & 1026.6 \\
$\nu_b$ & $e$       &&   463.6 & (262)$^a$&&354.3 &  280.9 &&  698.7 &  637.4 &&  703.6 &  687.9 \\ 
\hline \hline
\end{tabular}
\end{center}
\vspace*{1mm}
{\raggedright \hspace*{30mm} $^a$ \footnotesize Numerically not stable, see text.\\} 
\end{table*}
Concerning the CH$_3^+$--He frequencies, our results agree to within 10 cm$^{-1}$ with the computational results of Asvany and 
collaborators\cite{asvany1} -- except for the intermolecular coupling modes. Note that Asvany {\it et al.} used 2nd order 
vibrational perturbation theory, VPT2, based on a quartic force field obtained from conventional CCSD(T)/aug-cc-pVTZ 
calculations. This readily explains the differences between their and our calculations. 
A comparison with the experimental data shows that the CH stretching modes are nicely reproduced by 
our calculations, while the bending modes of the CH$_3^+$ moiety show larger deviations. 
Although the interaction of the helium
atom with the methyl cation is rather weak, its introduction leads to frequency shifts of more than 20 cm$^{-1}$ 
($\nu_2$) with respect to the bare methyl cation. \\

\begin{figure}
\includegraphics[width=\linewidth]{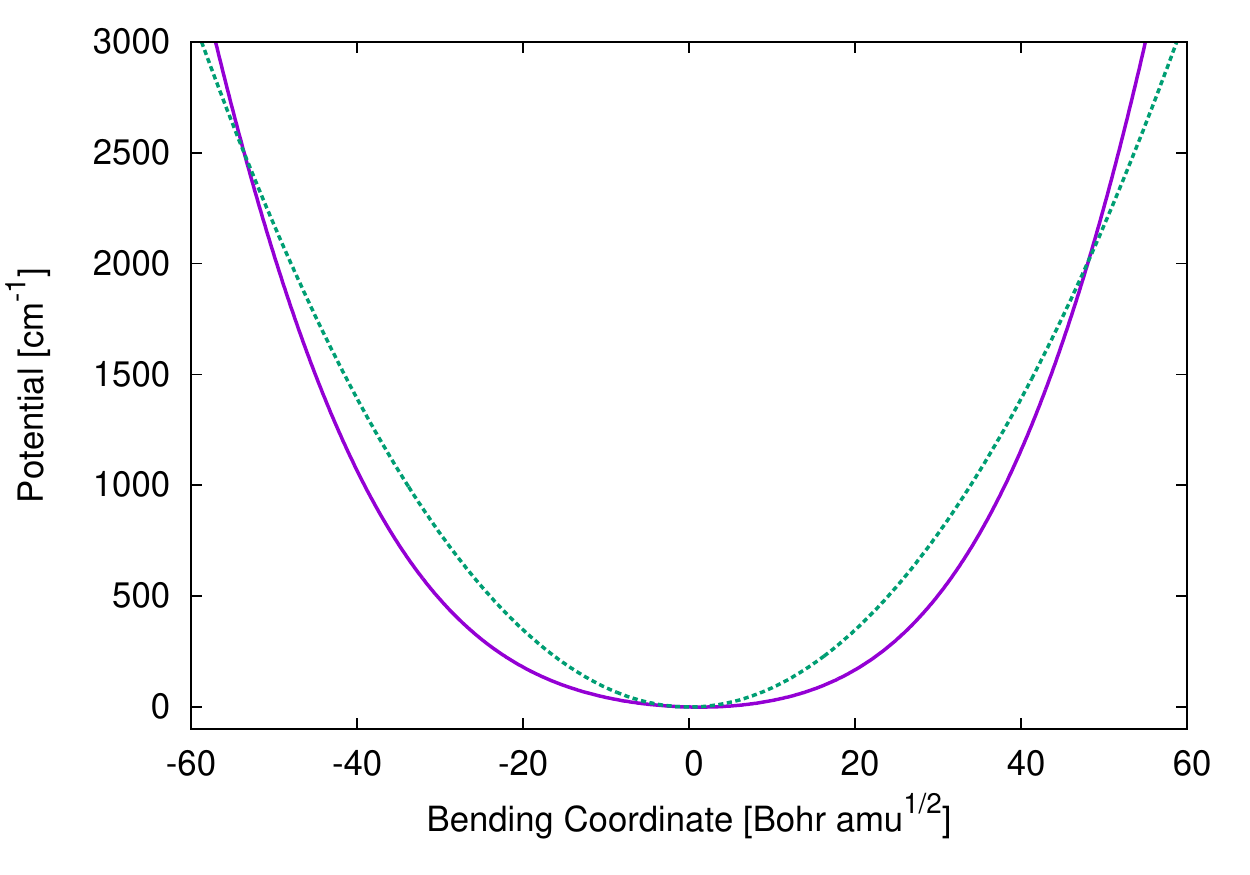}
\caption{Comparison of the harmonic potential (green, dotted) in comparison to the effective anharmonic VSCF potential
(purple, solid) along the intermolecular bending mode of CH$_3^+$--He.\label{potentials}}
\end{figure}

Of particular interest are the intermolecular bending modes $\nu_b$ of CH$_3^+$--He and CD$_3^+$--He. 
The most obvious and surprising effect 
concerning these modes are corrections due to anharmonicity of up to almost 50\%. 
With increasing size of the rare gas atom, the strong anharmonicity of the intermolecular bending mode
decreases and amounts to less than 3\% in case of the CH$_3^+$--Kr complex (cf.\ Table \ref{hfreq}).
The huge correction due to anharmonicity can already be seen
in the VSCF calculations and prompted us to visualize the harmonic potential in comparison to the 
effective anharmonic potential within the VSCF calculation of the vibrational ground state of CH$_3^+$--He, cf.\ Figure
\ref{potentials}.
Clearly, the effective potential shows a strong quartic contribution and thus the harmonic approximation
must be considered a rather poor approximation for this particular mode. \\

While all intramolecular modes of the CH$_3^+$ and CD$_3^+$ moieties and the intermolecular stretching modes $\nu_s$ 
were found to be stable with respect to the correlation space within the VCI calculations, 
the degenerate bending modes are very sensitive concerning the correlation space and thus the nature of the underlying VSCF modals. 
In principle, the VCI results 
should be independent of the nature of the one-mode wavefunctions employed in the configurations. However,
this was not the case for the intermolecular bending mode. The
VCI calculations for $\nu_b$ presented here, suffer from the fact, that the inclusion of high-lying correlation
functions did not improve the results, but introduced instabilities arising from the low lying dissociation
channel. Therefore, we had to limit the VCI space with respect to high excitations for the intermolecular stretching and 
the bending mode. As a results, the VCI calculations are not fully converged and do depend on the chosen basis. 
As must be expected, this problem vanishes for the clusters with larger values of D$_0$ and has not been observed for
the clusters containing Ar and Kr. In order to minimize the observed instabilities we investigated different 
sets of modals.
VCI calculations based on configurations built from a simple harmonic oscillator basis (without VSCF optimization) 
fail completely and yield results being significantly too high. On the 
other hand, standard state-specific VSCF modals lead to symmetry-broken solutions for these non-Abelian systems 
and are thus not a proper choice for these particular complexes. Therefore, we used modals obtained from ground-state 
VSCF calculations and from
configuration averaged VSCF calculations (CAVSCF) comprising the two configurations of the degenerate
bending modes. For the non-degenerate modes of course we did not use CAVSCF theory, but standard state-specific 
VSCF calculations. The resulting VCI frequencies for the intermolecular bending and stretching modes are 
summarized in Table \ref{trouble}.
\begin{table}
\caption{Comparison of the intermolecular coupling modes $\nu_b$ and $\nu_s$ obtained
from ground-state based VCI calculations (gs-VCI) and state-specific VCI calculations (ss-VCI).\label{trouble}}
\begin{center}
\begin{tabular}{lcrrcrr}
\hline \hline
        && \multicolumn{2}{c}{$\nu_b$} && \multicolumn{2}{c}{$\nu_s$} \\
        \cline{3-4} \cline{6-7} \\[-1ex]
Complex && gs-VCI & ss-VCI$^a$ && gs-VCI & ss-VCI$^b$ \\
\hline 
CH$_3^+$--He && 336.5 & 322.0 && 138.3 & 138.1 \\
CD$_3^+$--He && 269.8 & 261.5 && 142.0 & 141.8 \\
CH$_3^+$--Ne && 379.5 & 374.6 && 147.0 & 147.0 \\ 
CD$_3^+$--Ne && 283.3 & 280.9 && 141.2 & 141.1 \\
\hline \hline
\end{tabular}
\end{center}
\vspace*{1mm}
{\hspace*{5mm} $^a$ \footnotesize Using modals from CAVSCF theory.$\hfill$} \\
{\hspace*{5mm} $^b$ \footnotesize Using modals from state-specific VSCF calculations.$\hfill$} \\
\end{table}
According to Table \ref{trouble} the stretching modes $\nu_s$ are not affected by this problem and 
the ground-state based VCI calculations (gs-VCI) essentially coincide with the state-specific
VCI calculations (ss-VCI), which rely on CAVSCF modals for $\nu_b$ and standard state-specific VSCF modals for $\nu_s$.
Clearly, the instability is largest for CH$_3^+$--He and decreases with increasing values of D$_0$. For CD$_3^+$--Ne
the effect is in the range of the accuracy of our calculations and thus we consider this value to be trustworthy,
while the other three values bear a large error bar.
Note that, relying on VPT2 theory Asvany and co-workers\cite{asvany1} determined $\nu_b$ of CH$_3^+$--He 
to be 237 cm$^{-1}$ {\color{black}(cf.\ Table \ref{hfreq})}, which indicates even larger anharmonicity corrections. However, it is well known that VPT2 theory yields excellent
results for semi-rigid molecules, which are comparable to VCI results,\cite{chfclbr} but it 
remains an open question if 2nd order perturbation theory can handle floppy systems, whose potential energy surface cannot be 
described by simple quartic force fields and which lead to strong corrections like those observed for the systems here.
All other modes of the CH$_3^+$--He, CD$_3^+$--He and CH$_3^+$--Ne complexes were found to be converged 
with respect to the correlation space and thus coupling to the critical bending mode appears to be of minor importance.

\subsection{Tunneling splittings}
\begin{figure}
\includegraphics[width=\linewidth]{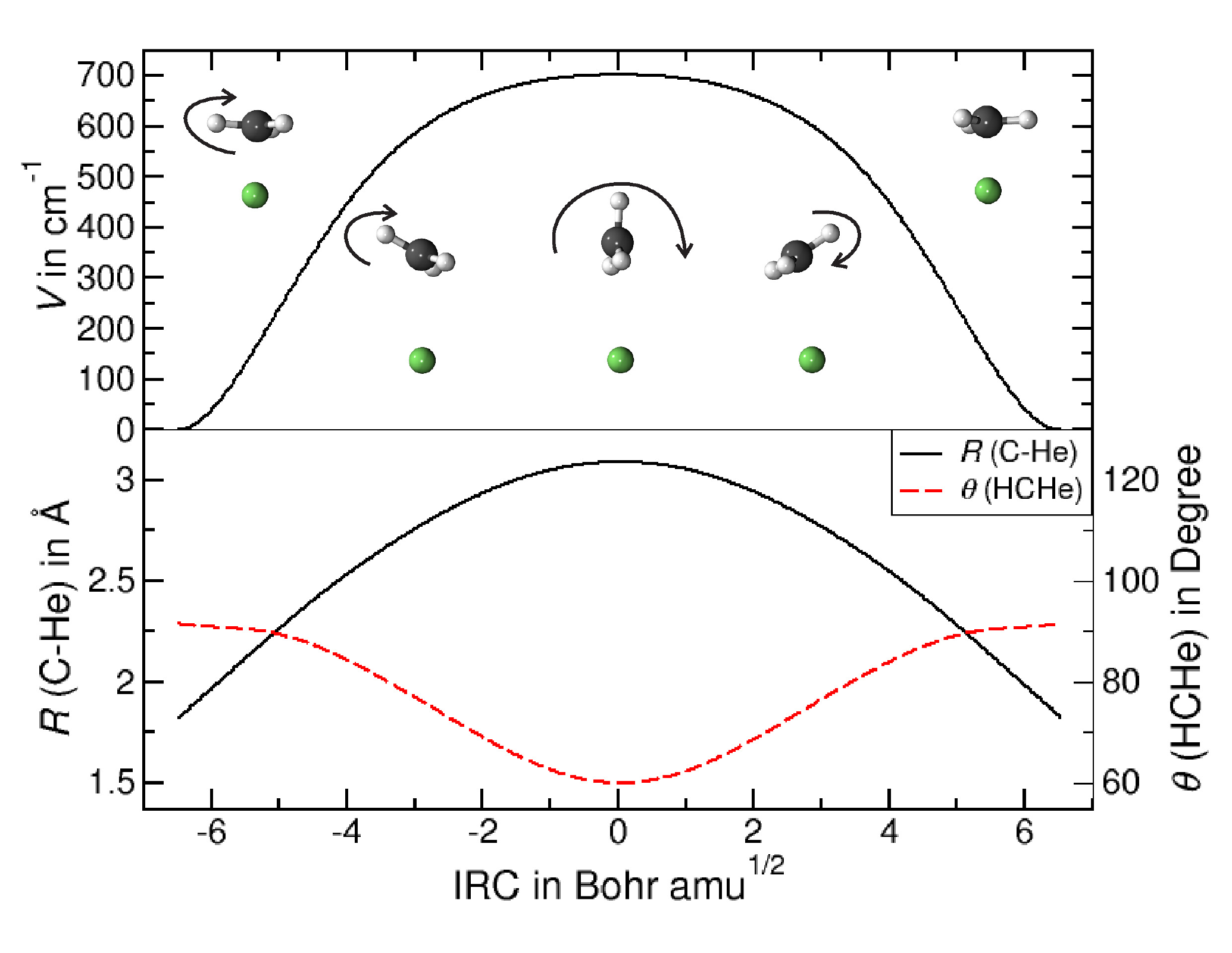}
\caption{Top: Potential energy along the minimum energy path for the intrinsic rotation of the methyl cation within the CH$_3^+$--He complex.
Bottom: 
Variation of the C--He distance $R$(C--He) (black solid) and the angle $\theta$(HCHe) (red dashed) along the same pathway.
\label{irc}}
\end{figure}

As mentioned above, all vibrational calculations were based on local PESs, which include just one of the
two 2p$_z$-bound minima. Attempts to generate larger PESs, which include both minima, failed due to the 
restriction of our program to rectilinear normal coordinates. In order to investigate the importance of 
tunneling, we performed several additional calculations. In a first step, we calculated the minimum energy
path along the intrinsic reaction coordinate between these minima. These calculations were started at the side-bound C$_{2v}$ transition
states, which was found to be 698.5 cm$^{-1}$ above the minima at the CCSD(T)-F12a/aug-cc-pVTZ level, but
about 127 cm$^{-1}$ below the dissociation energy ($D_\text{e}$).
Dopfer and co-workers determined the barrier to be slightly lower at about 600 cm$^{-1}$. \cite{luckhaus}
The potential energy along the intrinsic reaction coordinate (IRC) is shown in Figure~\ref{irc}.
The IRC consists of two principal movements: (1) Starting in one minimum of the PES, the distance between the helium and the carbon atoms will substantially be increased before (2) the methyl cation starts rotating as depicted in Figure~\ref{irc}. Note that the C--He distance, $R$(C--He) increases from 1.817~{\AA} in the minimum to
3.091~{\AA} in the transition state. The onset for the rotation appears to be rather late and, thus, the barrier for this reaction 
is fairly broad. For example, at about $s=-4$~Bohr~amu$^{1/2}$ the rotational angle 
$\theta$(HCHe) has changed by less than 10 degrees,
but the distance has already increased by more than 0.7 {\AA} and the energy has increased by about 400~cm$^{-1}$. At the beginning of the path, the barrier is largely controlled by the C--He distance rather than the movement of the hydrogen atoms. 

As the width of the potential has significant impact on the tunneling splitting, we studied 
this aspect in more detail and compared this IRC with that for the inversion reaction of the
hydronium cation, H$_3$O$^+$, which shows
very strong tunneling splittings of up to 373 cm$^{-1}$ for the fundamental modes.\cite{barrier_halonen,tunnel_halonen,tunnel_neff}
We believe that it is meaningful to compare these two systems as the barrier
height of both reactions is very similar and both reactions include three tunneling hydrogen atoms. According
to high-level calculations including different kinds of corrections, the barrier height for the H$_3$O$^+$ 
inversion was determined to be 651 or 657 cm$^{-1}$, respectively. For details see the 
work of Rajam\"aki {\it et al.}\cite{barrier_halonen,tunnel_halonen} and Neff {\it et al.}\cite{tunnel_neff} 
The comparison is shown in Figure~\ref{comparison}.
\begin{figure}
\includegraphics[width=\linewidth]{beide_IRCs-cm.pdf}
\caption{Comparison of the IRCs based on CCSD(T)-F12a calculations 
of inversions of CH$_3^+$--He (black) and the H$_3$O$^+$ (red dashed).\label{comparison}
Note, however, that the coordinates of the respective systems are obviously not the same.}
\end{figure}
The difference is striking. The width of the barrier for the hydronium cation is much smaller than for the 
CH$_3^+$--He complex. Consequently, one would expect much smaller tunneling splittings for the rare gas 
complexes studied here. Moreover, as the width does not decrease very much for higher energies, one would 
not expect significant tunneling splittings for vibrationally excited states. This expectation is in 
contrast to the large tunneling splittings as predicted by the calculations of Dopfer and Luckhaus.\cite{luckhaus}
Of course, the presented IRC is no proof that tunneling may not occur, but 
it is an indication that it is probably of little importance. This would be in agreement of the recent observations
of T{\"o}pfer {\it et al.} \cite{asvany2} \\


The tunneling splitting of the vibrational ground state
of the CH$_3^+$--He complex was determined by instanton calculations. As outlined in the Computational
Details, these calculations were not restricted to the local potential, which has been employed for 
the VCI calculations, but used the same electronic structure level, i.e.\ CCSD(T)-F12a/aug-cc-pVTZ, on the fly. The instanton
calculations predict the tunneling splitting for the vibrational ground state to be less than $10^{-4}$ cm$^{-1}$
and, thus, to be at least one to two orders of magnitude smaller than the value determined by Dopfer and 
Luckhaus of $8\cdot 10^{-3}$ cm$^{-1}$.\cite{luckhaus} Our result takes into account the 3-fold degeneracy 
of the path in the partition function, which is caused by 
the three equivalent tunneling paths arising from the C$_{3v}$ point group of the system. 
Note that semiclassical instanton theory has an intrinsic error bar, which is difficult to estimate. However,
Richardson\cite{richardson} computed the tunneling splitting of the formic acid dimer with ring-polymer instanton theory
and observed an error being not larger than 20\% in comparison to accurate quantum-dynamical 
calculations. Moreover, he concluded that it appears to be important
to include all dimensions within tunneling calculations 
and to use an accurate PES (as we have done in our calculations) rather than
using reduced dimensionality approaches. \cite{beyer2016,richardson2018}
Therefore, we consider a value of $10^{-4}$ cm$^{-1}$ to be 
an estimation of an upper bound, {\color{black} which we consider to be more reliable than
the value of Dopfer and Luckhaus, which was obtained from calculations restricted
to a frozen CH$_3^+$ moiety and to an MP2 potential of the three intermolecular coordinates only.}

Again, this is no proof that significant tunneling splitting may not show up for excited vibrational 
states, which we could not compute directly, but it makes it rather unlikely. For that reason, we did not
perform comparable calculations for the complexes with heavier rare gas atoms, in which tunneling processes 
are even more unlikely due to the higher mass of the system and the pronounced covalent character of the 
interaction.

\section{Summary and conclusions}
High-level {\it ab initio} methods have been used to study the vibrational spectra of CH$_3^+$--Rg 
(Rg=He, Ne, Ar, Kr) complexes.
These calculations largely confirm the experimental results of Dopfer and co-workers 
{\color{black} and the conclusions of these authors that the bonding mechanism is controlled
by induction for the two smaller rare gas atoms, while a covalent bond with a pronounced
charge transfer can be seen for the Ar and Kr-complexes.} In extension 
to previous studies, high-level predictions have been provided for all fundamental vibrational transitions 
of these complexes. These were based on variational VCI calculations relying on multidimensional
potential energy surfaces obtained from explicitly correlated coupled-cluster theory. Results for 
the {\color{black} vibrational frequencies} of the CH$_3^+$--Kr complex have been provided for 
the first time. \\

The intermolecular bending modes of the CH$_3^+$--He complex could not be determined with high accuracy due
to numerical instabilities arising from the very low lying dissociation channel of the complex. The most 
reliable value was obtained from configuration-averaged vibrational self-consistent field theory, which was 
introduced to optimize the modal basis prior to the VCI calculations for degenerate states.\\

Reference calculations for the bare methyl cation show that corrections due to core-valence correlation,
high-level coupled-cluster terms, scalar-relativistic effects and terms of higher order within
the expansion of the $\mu$-tensor are of minor importance. Consequently, standard CCSD(T)-F12b/aug-cc-pVTZ
calculations already provide excellent estimates at much lower computational cost.\\

The aspect of tunneling splittings has been studied in detail. While tunneling splitting for
vibrationally excited states could not be computed directly for technical reasons, instanton
theory provides a splitting for the vibrational ground-state, which is much smaller than
obtained from previous calculations. This and the very broad barrier of the potential indicates
that tunneling effects might be negligible even for vibrationally excited states.

\section{Acknowledgments}
We are grateful to Dr.\ O.\ Asvany and B.\ Ziegler for valuable and insightful discussions.
Moreover, we thank Prof.\ A.\ K\"ohn for helping us to perform CCSDT and CCSDTQ calculation using the GeCCo 
suite of programs.
Support was provided by the COST Action CM1405 "Molecules in Motion (MOLIM)".
This research was also supported in part by the bwHPC initiative and the bwHPC-C5 project
provided through associated compute services of the JUSTUS HPC facility at the University
of Ulm. bwHPC and bwHPC-C5 are funded by the Ministry of Science, Research and the Arts
Baden-W\"urttemberg (MWK). Further financial support was provided by
the Deutsche Forschungsgemeinschaft (DFG), project RA 656/25-1.
\begin{center}
\end{center}


\begin{thebibliography}{62}
\expandafter\ifx\csname natexlab\endcsname\relax\def\natexlab#1{#1}\fi
\expandafter\ifx\csname bibnamefont\endcsname\relax
  \def\bibnamefont#1{#1}\fi
\expandafter\ifx\csname bibfnamefont\endcsname\relax
  \def\bibfnamefont#1{#1}\fi
\expandafter\ifx\csname citenamefont\endcsname\relax
  \def\citenamefont#1{#1}\fi
\expandafter\ifx\csname url\endcsname\relax
  \def\url#1{\texttt{#1}}\fi
\expandafter\ifx\csname urlprefix\endcsname\relax\def\urlprefix{URL }\fi
\providecommand{\bibinfo}[2]{#2}
\providecommand{\eprint}[2][]{\url{#2}}

\bibitem[{\citenamefont{Olkhov et~al.}(1999)\citenamefont{Olkhov, Nizkorodov,
  and Dopfer}}]{hech3}
\bibinfo{author}{\bibfnamefont{R.~V.} \bibnamefont{Olkhov}},
  \bibinfo{author}{\bibfnamefont{S.~A.} \bibnamefont{Nizkorodov}},
  \bibnamefont{and} \bibinfo{author}{\bibfnamefont{O.}~\bibnamefont{Dopfer}},
  \bibinfo{journal}{J.~Chem.~Phys.} \textbf{\bibinfo{volume}{110}},
  \bibinfo{pages}{9527} (\bibinfo{year}{1999}).

\bibitem[{\citenamefont{Dopfer et~al.}(2000)\citenamefont{Dopfer, Olkhov, and
  Maier}}]{nech3}
\bibinfo{author}{\bibfnamefont{O.}~\bibnamefont{Dopfer}},
  \bibinfo{author}{\bibfnamefont{R.~V.} \bibnamefont{Olkhov}},
  \bibnamefont{and} \bibinfo{author}{\bibfnamefont{J.~P.} \bibnamefont{Maier}},
  \bibinfo{journal}{J.~Chem.~Phys.} \textbf{\bibinfo{volume}{112}},
  \bibinfo{pages}{2176} (\bibinfo{year}{2000}).

\bibitem[{\citenamefont{Olkhov et~al.}(1998)\citenamefont{Olkhov, Nizkorodov,
  and Dopfer}}]{arch3}
\bibinfo{author}{\bibfnamefont{R.~V.} \bibnamefont{Olkhov}},
  \bibinfo{author}{\bibfnamefont{S.~A.} \bibnamefont{Nizkorodov}},
  \bibnamefont{and} \bibinfo{author}{\bibfnamefont{O.}~\bibnamefont{Dopfer}},
  \bibinfo{journal}{J.~Chem.~Phys.} \textbf{\bibinfo{volume}{108}},
  \bibinfo{pages}{10046} (\bibinfo{year}{1998}).

\bibitem[{\citenamefont{Dopfer and Luckhaus}(2002)}]{luckhaus}
\bibinfo{author}{\bibfnamefont{O.}~\bibnamefont{Dopfer}} \bibnamefont{and}
  \bibinfo{author}{\bibfnamefont{D.}~\bibnamefont{Luckhaus}},
  \bibinfo{journal}{J.~Chem.~Phys.} \textbf{\bibinfo{volume}{116}},
  \bibinfo{pages}{1012} (\bibinfo{year}{2002}).

\bibitem[{\citenamefont{Dopfer}(2003)}]{dopfer_review}
\bibinfo{author}{\bibfnamefont{O.}~\bibnamefont{Dopfer}},
  \bibinfo{journal}{Int.\ Rev.\ Phys.\ Chem.} \textbf{\bibinfo{volume}{22}},
  \bibinfo{pages}{437} (\bibinfo{year}{2003}).

\bibitem[{\citenamefont{Asvany et~al.}(2018)\citenamefont{Asvany, Thorwirth,
  Redlich, and Schlemmer}}]{asvany1}
\bibinfo{author}{\bibfnamefont{O.}~\bibnamefont{Asvany}},
  \bibinfo{author}{\bibfnamefont{S.}~\bibnamefont{Thorwirth}},
  \bibinfo{author}{\bibfnamefont{B.}~\bibnamefont{Redlich}}, \bibnamefont{and}
  \bibinfo{author}{\bibfnamefont{S.}~\bibnamefont{Schlemmer}},
  \bibinfo{journal}{J. Mol. Spectrosc.} \textbf{\bibinfo{volume}{347}},
  \bibinfo{pages}{1} (\bibinfo{year}{2018}).

\bibitem[{\citenamefont{Hiraoka and Kudaka}(1991)}]{hiraoka}
\bibinfo{author}{\bibfnamefont{K.}~\bibnamefont{Hiraoka}} \bibnamefont{and}
  \bibinfo{author}{\bibfnamefont{I.}~\bibnamefont{Kudaka}},
  \bibinfo{journal}{Chem.~Phys.~Lett.} \textbf{\bibinfo{volume}{178}},
  \bibinfo{pages}{103} (\bibinfo{year}{1991}).

\bibitem[{\citenamefont{{D. A.~Clabo~Jr. and W. D.~Allen and R. B.~Remington
  and Y.~Yamaguchi and H. F.~Schaefer}}(1988)}]{vpt2}
\bibinfo{author}{\bibnamefont{{D. A.~Clabo~Jr. and W. D.~Allen and R.
  B.~Remington and Y.~Yamaguchi and H. F.~Schaefer}}},
  \bibinfo{journal}{Chem.~Phys.} \textbf{\bibinfo{volume}{123}},
  \bibinfo{pages}{187} (\bibinfo{year}{1988}).

\bibitem[{\citenamefont{T{\"o}pfer et~al.}(2018)\citenamefont{T{\"o}pfer,
  Salomon, Schlemmer, Asvany, Dopfer, Kohguchi, and Yamada}}]{asvany2}
\bibinfo{author}{\bibfnamefont{M.}~\bibnamefont{T{\"o}pfer}},
  \bibinfo{author}{\bibfnamefont{T.}~\bibnamefont{Salomon}},
  \bibinfo{author}{\bibfnamefont{S.}~\bibnamefont{Schlemmer}},
  \bibinfo{author}{\bibfnamefont{O.}~\bibnamefont{Asvany}},
  \bibinfo{author}{\bibfnamefont{O.}~\bibnamefont{Dopfer}},
  \bibinfo{author}{\bibfnamefont{H.}~\bibnamefont{Kohguchi}}, \bibnamefont{and}
  \bibinfo{author}{\bibfnamefont{K.~M.~T.} \bibnamefont{Yamada}},
  \bibinfo{journal}{Phys. Rev. Lett.} \textbf{\bibinfo{volume}{121}},
  \bibinfo{pages}{143001} (\bibinfo{year}{2018}).

\bibitem[{\citenamefont{Zerner}(1989)}]{zerner}
\bibinfo{author}{\bibfnamefont{M.~C.} \bibnamefont{Zerner}},
  \bibinfo{journal}{Int.~J.~Quantum Chem.} \textbf{\bibinfo{volume}{35}},
  \bibinfo{pages}{567} (\bibinfo{year}{1989}).

\bibitem[{\citenamefont{van~den Heuvel et~al.}(2016)\citenamefont{van~den
  Heuvel, Calvello, and Soncini}}]{soncini}
\bibinfo{author}{\bibfnamefont{W.}~\bibnamefont{van~den Heuvel}},
  \bibinfo{author}{\bibfnamefont{S.}~\bibnamefont{Calvello}}, \bibnamefont{and}
  \bibinfo{author}{\bibfnamefont{A.}~\bibnamefont{Soncini}},
  \bibinfo{journal}{Phys.~Chem.~Chem.~Phys.} \textbf{\bibinfo{volume}{18}},
  \bibinfo{pages}{15807} (\bibinfo{year}{2016}).

\bibitem[{\citenamefont{Hallmen et~al.}(2017)\citenamefont{Hallmen, K{\"o}ppl,
  Rauhut, Stoll, and van Slageren}}]{phil1}
\bibinfo{author}{\bibfnamefont{P.~P.} \bibnamefont{Hallmen}},
  \bibinfo{author}{\bibfnamefont{C.}~\bibnamefont{K{\"o}ppl}},
  \bibinfo{author}{\bibfnamefont{G.}~\bibnamefont{Rauhut}},
  \bibinfo{author}{\bibfnamefont{H.}~\bibnamefont{Stoll}}, \bibnamefont{and}
  \bibinfo{author}{\bibfnamefont{J.}~\bibnamefont{van Slageren}},
  \bibinfo{journal}{J.~Chem.~Phys.} \textbf{\bibinfo{volume}{147}},
  \bibinfo{pages}{164101} (\bibinfo{year}{2017}).

\bibitem[{\citenamefont{Neff et~al.}(2011)\citenamefont{Neff, Hrenar,
  Oschetzki, and Rauhut}}]{convergence}
\bibinfo{author}{\bibfnamefont{M.}~\bibnamefont{Neff}},
  \bibinfo{author}{\bibfnamefont{T.}~\bibnamefont{Hrenar}},
  \bibinfo{author}{\bibfnamefont{D.}~\bibnamefont{Oschetzki}},
  \bibnamefont{and} \bibinfo{author}{\bibfnamefont{G.}~\bibnamefont{Rauhut}},
  \bibinfo{journal}{J.~Chem.~Phys.} \textbf{\bibinfo{volume}{134}},
  \bibinfo{pages}{064105} (\bibinfo{year}{2011}).

\bibitem[{\citenamefont{Watson}(1968)}]{watson}
\bibinfo{author}{\bibfnamefont{J.~K.~G.} \bibnamefont{Watson}},
  \bibinfo{journal}{Mol.~Phys.} \textbf{\bibinfo{volume}{15}},
  \bibinfo{pages}{479} (\bibinfo{year}{1968}).

\bibitem[{\citenamefont{{J. M. Bowman and T. Carrington Jr. and H. D.
  Meyer}}(2008)}]{vreview}
\bibinfo{author}{\bibnamefont{{J. M. Bowman and T. Carrington Jr. and H. D.
  Meyer}}}, \bibinfo{journal}{Mol.~Phys.} \textbf{\bibinfo{volume}{106}},
  \bibinfo{pages}{2145} (\bibinfo{year}{2008}).

\bibitem[{\citenamefont{Ziegler and Rauhut}(2016)}]{fit}
\bibinfo{author}{\bibfnamefont{B.}~\bibnamefont{Ziegler}} \bibnamefont{and}
  \bibinfo{author}{\bibfnamefont{G.}~\bibnamefont{Rauhut}},
  \bibinfo{journal}{J. Chem. Phys.} \textbf{\bibinfo{volume}{144}},
  \bibinfo{pages}{114114} (\bibinfo{year}{2016}).

\bibitem[{\citenamefont{Bowman}(1986)}]{bowman1}
\bibinfo{author}{\bibfnamefont{J.~M.} \bibnamefont{Bowman}},
  \bibinfo{journal}{Acc.~Chem.~Res.} \textbf{\bibinfo{volume}{19}},
  \bibinfo{pages}{202} (\bibinfo{year}{1986}).

\bibitem[{\citenamefont{Bowman}(1978)}]{bowman2}
\bibinfo{author}{\bibfnamefont{J.~M.} \bibnamefont{Bowman}},
  \bibinfo{journal}{J.~Chem.~Phys.} \textbf{\bibinfo{volume}{68}},
  \bibinfo{pages}{608} (\bibinfo{year}{1978}).

\bibitem[{\citenamefont{Gerber and Ratner}(1988)}]{gerber1}
\bibinfo{author}{\bibfnamefont{R.}~\bibnamefont{Gerber}} \bibnamefont{and}
  \bibinfo{author}{\bibfnamefont{M.}~\bibnamefont{Ratner}},
  \bibinfo{journal}{Adv.~Chem.~Phys.} \textbf{\bibinfo{volume}{70}},
  \bibinfo{pages}{97} (\bibinfo{year}{1988}).

\bibitem[{\citenamefont{Miller}(1975)}]{mil75}
\bibinfo{author}{\bibfnamefont{W.~H.} \bibnamefont{Miller}},
  \bibinfo{journal}{J. Chem. Phys.} \textbf{\bibinfo{volume}{62}},
  \bibinfo{pages}{1899} (\bibinfo{year}{1975}).

\bibitem[{\citenamefont{Va{\u \i}nshte{\u \i}n et~al.}(1982)\citenamefont{Va{\u
  \i}nshte{\u \i}n, Zakharov, Novikov, and Shifman}}]{vai82}
\bibinfo{author}{\bibfnamefont{A.~I.} \bibnamefont{Va{\u \i}nshte{\u \i}n}},
  \bibinfo{author}{\bibfnamefont{V.~I.} \bibnamefont{Zakharov}},
  \bibinfo{author}{\bibfnamefont{V.~A.} \bibnamefont{Novikov}},
  \bibnamefont{and} \bibinfo{author}{\bibfnamefont{M.~A.}
  \bibnamefont{Shifman}}, \bibinfo{journal}{Sov. Phys. Usp.}
  \textbf{\bibinfo{volume}{25}}, \bibinfo{pages}{195} (\bibinfo{year}{1982}).

\bibitem[{\citenamefont{Mil'nikov and Nakamura}(2008)}]{mil08a}
\bibinfo{author}{\bibfnamefont{G.}~\bibnamefont{Mil'nikov}} \bibnamefont{and}
  \bibinfo{author}{\bibfnamefont{H.}~\bibnamefont{Nakamura}},
  \bibinfo{journal}{Phys. Chem. Chem. Phys.} \textbf{\bibinfo{volume}{10}},
  \bibinfo{pages}{1374} (\bibinfo{year}{2008}).

\bibitem[{\citenamefont{Richardson and Althorpe}(2011)}]{ric11}
\bibinfo{author}{\bibfnamefont{J.~O.} \bibnamefont{Richardson}}
  \bibnamefont{and} \bibinfo{author}{\bibfnamefont{S.~C.}
  \bibnamefont{Althorpe}}, \bibinfo{journal}{J. Chem. Phys.}
  \textbf{\bibinfo{volume}{134}}, \bibinfo{pages}{054109}
  (\bibinfo{year}{2011}).

\bibitem[{\citenamefont{Richardson et~al.}(2011)\citenamefont{Richardson,
  Althorpe, and Wales}}]{richardson2011}
\bibinfo{author}{\bibfnamefont{J.~O.} \bibnamefont{Richardson}},
  \bibinfo{author}{\bibfnamefont{S.~C.} \bibnamefont{Althorpe}},
  \bibnamefont{and} \bibinfo{author}{\bibfnamefont{D.~J.} \bibnamefont{Wales}},
  \bibinfo{journal}{J. Chem. Phys.} \textbf{\bibinfo{volume}{135}},
  \bibinfo{pages}{124109} (\bibinfo{year}{2011}).

\bibitem[{\citenamefont{Adler et~al.}(2007)\citenamefont{Adler, Knizia, and
  Werner}}]{ccf12}
\bibinfo{author}{\bibfnamefont{T.~B.} \bibnamefont{Adler}},
  \bibinfo{author}{\bibfnamefont{G.}~\bibnamefont{Knizia}}, \bibnamefont{and}
  \bibinfo{author}{\bibfnamefont{H.-J.} \bibnamefont{Werner}},
  \bibinfo{journal}{J.~Chem.~Phys.} \textbf{\bibinfo{volume}{127}},
  \bibinfo{pages}{221106} (\bibinfo{year}{2007}).

\bibitem[{\citenamefont{Woon and Jr}(1993)}]{vtz}
\bibinfo{author}{\bibfnamefont{D.~E.} \bibnamefont{Woon}} \bibnamefont{and}
  \bibinfo{author}{\bibfnamefont{T.~H.~D.} \bibnamefont{Jr}},
  \bibinfo{journal}{J.~Chem.~Phys.} \textbf{\bibinfo{volume}{98}},
  \bibinfo{pages}{1358} (\bibinfo{year}{1993}).

\bibitem[{\citenamefont{Rauhut et~al.}(2009)\citenamefont{Rauhut, Knizia, and
  Werner}}]{knizia09}
\bibinfo{author}{\bibfnamefont{G.}~\bibnamefont{Rauhut}},
  \bibinfo{author}{\bibfnamefont{G.}~\bibnamefont{Knizia}}, \bibnamefont{and}
  \bibinfo{author}{\bibfnamefont{H.}~\bibnamefont{Werner}},
  \bibinfo{journal}{J.~Chem.~Phys.} \textbf{\bibinfo{volume}{130}},
  \bibinfo{pages}{054105} (\bibinfo{year}{2009}).

\bibitem[{\citenamefont{Werner et~al.}(2018)\citenamefont{Werner, Knowles,
  Lindh, Manby, {Sch\"{u}tz} et~al.}}]{MOLPRO_dev_brief}
\bibinfo{author}{\bibfnamefont{H.-J.} \bibnamefont{Werner}},
  \bibinfo{author}{\bibfnamefont{P.~J.} \bibnamefont{Knowles}},
  \bibinfo{author}{\bibfnamefont{R.}~\bibnamefont{Lindh}},
  \bibinfo{author}{\bibfnamefont{F.~R.} \bibnamefont{Manby}},
  \bibinfo{author}{\bibfnamefont{M.}~\bibnamefont{{Sch\"{u}tz}}},
  \bibnamefont{et~al.}, \emph{\bibinfo{title}{Molpro, development version
  2018.2, a package of ab initio programs}} (\bibinfo{year}{2018}),
  \bibinfo{note}{see http://www.molpro.net}.

\bibitem[{\citenamefont{Neff and Rauhut}(2009{\natexlab{a}})}]{neff1}
\bibinfo{author}{\bibfnamefont{M.}~\bibnamefont{Neff}} \bibnamefont{and}
  \bibinfo{author}{\bibfnamefont{G.}~\bibnamefont{Rauhut}},
  \bibinfo{journal}{J.~Chem.~Phys.} \textbf{\bibinfo{volume}{131}},
  \bibinfo{pages}{124129} (\bibinfo{year}{2009}{\natexlab{a}}).

\bibitem[{\citenamefont{Neff and Rauhut}(2009{\natexlab{b}})}]{neff2}
\bibinfo{author}{\bibfnamefont{M.}~\bibnamefont{Neff}} \bibnamefont{and}
  \bibinfo{author}{\bibfnamefont{G.}~\bibnamefont{Rauhut}},
  \bibinfo{journal}{J.~Chem.~Phys.} \textbf{\bibinfo{volume}{131}},
  \bibinfo{pages}{229901} (\bibinfo{year}{2009}{\natexlab{b}}).

\bibitem[{\citenamefont{Ziegler and Rauhut}(2018)}]{benni}
\bibinfo{author}{\bibfnamefont{B.}~\bibnamefont{Ziegler}} \bibnamefont{and}
  \bibinfo{author}{\bibfnamefont{G.}~\bibnamefont{Rauhut}},
  \bibinfo{journal}{J.~Chem.~Phys.} \textbf{\bibinfo{volume}{149}},
  \bibinfo{pages}{164110} (\bibinfo{year}{2018}).

\bibitem[{\citenamefont{Rommel et~al.}(2011)\citenamefont{Rommel, Goumans, and
  K\"astner}}]{rommel2011}
\bibinfo{author}{\bibfnamefont{J.~B.} \bibnamefont{Rommel}},
  \bibinfo{author}{\bibfnamefont{T.~P.~M.} \bibnamefont{Goumans}},
  \bibnamefont{and}
  \bibinfo{author}{\bibfnamefont{J.}~\bibnamefont{K\"astner}},
  \bibinfo{journal}{J. Chem. Theory Comput.} \textbf{\bibinfo{volume}{7}},
  \bibinfo{pages}{690} (\bibinfo{year}{2011}).

\bibitem[{\citenamefont{K\"astner et~al.}(2009)\citenamefont{K\"astner, Carr,
  Keal, Thiel, Wander, and Sherwood}}]{kaestner2009}
\bibinfo{author}{\bibfnamefont{J.}~\bibnamefont{K\"astner}},
  \bibinfo{author}{\bibfnamefont{J.~M.} \bibnamefont{Carr}},
  \bibinfo{author}{\bibfnamefont{T.~W.} \bibnamefont{Keal}},
  \bibinfo{author}{\bibfnamefont{W.}~\bibnamefont{Thiel}},
  \bibinfo{author}{\bibfnamefont{A.}~\bibnamefont{Wander}}, \bibnamefont{and}
  \bibinfo{author}{\bibfnamefont{P.}~\bibnamefont{Sherwood}},
  \bibinfo{journal}{J. Phys. Chem. A} \textbf{\bibinfo{volume}{113}},
  \bibinfo{pages}{11856} (\bibinfo{year}{2009}).

\bibitem[{\citenamefont{Metz et~al.}(2014)\citenamefont{Metz, K\"astner, Sokol,
  Keal, and Sherwood}}]{metz2014}
\bibinfo{author}{\bibfnamefont{S.}~\bibnamefont{Metz}},
  \bibinfo{author}{\bibfnamefont{J.}~\bibnamefont{K\"astner}},
  \bibinfo{author}{\bibfnamefont{A.~A.} \bibnamefont{Sokol}},
  \bibinfo{author}{\bibfnamefont{T.~W.} \bibnamefont{Keal}}, \bibnamefont{and}
  \bibinfo{author}{\bibfnamefont{P.}~\bibnamefont{Sherwood}},
  \bibinfo{journal}{WIREs Comput. Mol. Sci.} \textbf{\bibinfo{volume}{4}},
  \bibinfo{pages}{101} (\bibinfo{year}{2014}).

\bibitem[{\citenamefont{Hratchian and Schlegel}(2004)}]{hratchian2004}
\bibinfo{author}{\bibfnamefont{H.~P.} \bibnamefont{Hratchian}}
  \bibnamefont{and} \bibinfo{author}{\bibfnamefont{H.~B.}
  \bibnamefont{Schlegel}}, \bibinfo{journal}{J.Chem. Phys.}
  \textbf{\bibinfo{volume}{120}}, \bibinfo{pages}{9918} (\bibinfo{year}{2004}).

\bibitem[{\citenamefont{Hratchian and Schlegel}(2005)}]{hratchian2005}
\bibinfo{author}{\bibfnamefont{H.~P.} \bibnamefont{Hratchian}}
  \bibnamefont{and} \bibinfo{author}{\bibfnamefont{H.~B.}
  \bibnamefont{Schlegel}}, \bibinfo{journal}{J. Chem. Theory Comput.}
  \textbf{\bibinfo{volume}{1}}, \bibinfo{pages}{61} (\bibinfo{year}{2005}).

\bibitem[{\citenamefont{Meisner et~al.}(2017)\citenamefont{Meisner, Markmeyer,
  Bohner, and K\"astner}}]{meisner2017b}
\bibinfo{author}{\bibfnamefont{J.}~\bibnamefont{Meisner}},
  \bibinfo{author}{\bibfnamefont{M.~N.} \bibnamefont{Markmeyer}},
  \bibinfo{author}{\bibfnamefont{M.~U.} \bibnamefont{Bohner}},
  \bibnamefont{and}
  \bibinfo{author}{\bibfnamefont{J.}~\bibnamefont{K\"astner}},
  \bibinfo{journal}{Phys. Chem. Chem. Phys.} \textbf{\bibinfo{volume}{19}},
  \bibinfo{pages}{23085} (\bibinfo{year}{2017}).

\bibitem[{\citenamefont{Reed et~al.}(1985)\citenamefont{Reed, Weinstock, and
  Weinhold}}]{nbo}
\bibinfo{author}{\bibfnamefont{A.~E.} \bibnamefont{Reed}},
  \bibinfo{author}{\bibfnamefont{R.~B.} \bibnamefont{Weinstock}},
  \bibnamefont{and} \bibinfo{author}{\bibfnamefont{F.}~\bibnamefont{Weinhold}},
  \bibinfo{journal}{J.~Chem.~Phys.} \textbf{\bibinfo{volume}{83}},
  \bibinfo{pages}{735} (\bibinfo{year}{1985}).

\bibitem[{\citenamefont{Hesselmann et~al.}(2005)\citenamefont{Hesselmann,
  Jansen, and Sch{\"u}tz}}]{sapt}
\bibinfo{author}{\bibfnamefont{A.}~\bibnamefont{Hesselmann}},
  \bibinfo{author}{\bibfnamefont{G.}~\bibnamefont{Jansen}}, \bibnamefont{and}
  \bibinfo{author}{\bibfnamefont{M.}~\bibnamefont{Sch{\"u}tz}},
  \bibinfo{journal}{J.~Chem.~Phys.} \textbf{\bibinfo{volume}{122}},
  \bibinfo{pages}{014103} (\bibinfo{year}{2005}).

\bibitem[{\citenamefont{Crofton et~al.}(1988)\citenamefont{Crofton, Jagod,
  Rehfuss, Kreiner, and Oka}}]{crofton}
\bibinfo{author}{\bibfnamefont{M.~W.} \bibnamefont{Crofton}},
  \bibinfo{author}{\bibfnamefont{M.-F.} \bibnamefont{Jagod}},
  \bibinfo{author}{\bibfnamefont{B.~D.} \bibnamefont{Rehfuss}},
  \bibinfo{author}{\bibfnamefont{W.~A.} \bibnamefont{Kreiner}},
  \bibnamefont{and} \bibinfo{author}{\bibfnamefont{T.}~\bibnamefont{Oka}},
  \bibinfo{journal}{J.~Chem.~Phys.} \textbf{\bibinfo{volume}{88}},
  \bibinfo{pages}{666} (\bibinfo{year}{1988}).

\bibitem[{\citenamefont{Koenig et~al.}(1975)\citenamefont{Koenig, Balle, and
  Snell}}]{koenig1}
\bibinfo{author}{\bibfnamefont{T.}~\bibnamefont{Koenig}},
  \bibinfo{author}{\bibfnamefont{T.}~\bibnamefont{Balle}}, \bibnamefont{and}
  \bibinfo{author}{\bibfnamefont{W.}~\bibnamefont{Snell}},
  \bibinfo{journal}{J.~Am.~Chem.~Soc.} \textbf{\bibinfo{volume}{97}},
  \bibinfo{pages}{662} (\bibinfo{year}{1975}).

\bibitem[{\citenamefont{Koenig et~al.}(1976)\citenamefont{Koenig, Balle, and
  Chang}}]{koenig2}
\bibinfo{author}{\bibfnamefont{T.}~\bibnamefont{Koenig}},
  \bibinfo{author}{\bibfnamefont{T.}~\bibnamefont{Balle}}, \bibnamefont{and}
  \bibinfo{author}{\bibfnamefont{J.~C.} \bibnamefont{Chang}},
  \bibinfo{journal}{Spectrosc. Lett.} \textbf{\bibinfo{volume}{9}},
  \bibinfo{pages}{755} (\bibinfo{year}{1976}).

\bibitem[{\citenamefont{Dyke et~al.}(1976)\citenamefont{Dyke, Jonathan, Lee,
  and Morris}}]{dyke}
\bibinfo{author}{\bibfnamefont{J.}~\bibnamefont{Dyke}},
  \bibinfo{author}{\bibfnamefont{N.}~\bibnamefont{Jonathan}},
  \bibinfo{author}{\bibfnamefont{E.}~\bibnamefont{Lee}}, \bibnamefont{and}
  \bibinfo{author}{\bibfnamefont{A.}~\bibnamefont{Morris}},
  \bibinfo{journal}{J. Chem. Soc., Faraday Trans. II}
  \textbf{\bibinfo{volume}{72}}, \bibinfo{pages}{1385} (\bibinfo{year}{1976}).

\bibitem[{\citenamefont{Liu et~al.}(2001)\citenamefont{Liu, Gross, and
  Suits}}]{liu}
\bibinfo{author}{\bibfnamefont{X.}~\bibnamefont{Liu}},
  \bibinfo{author}{\bibfnamefont{R.~L.} \bibnamefont{Gross}}, \bibnamefont{and}
  \bibinfo{author}{\bibfnamefont{A.~G.} \bibnamefont{Suits}},
  \bibinfo{journal}{Science} \textbf{\bibinfo{volume}{294}},
  \bibinfo{pages}{2527} (\bibinfo{year}{2001}).

\bibitem[{\citenamefont{{Cunha~de~Miranda}
  et~al.}(2010)\citenamefont{{Cunha~de~Miranda}, Acaraz, Elhanine, Noller,
  Hemberger, Fischer, Garcia, Soldi-Lose, Gans, {Viera Mendes}
  et~al.}}]{pbotsch}
\bibinfo{author}{\bibfnamefont{B.~K.} \bibnamefont{{Cunha~de~Miranda}}},
  \bibinfo{author}{\bibfnamefont{C.}~\bibnamefont{Acaraz}},
  \bibinfo{author}{\bibfnamefont{M.}~\bibnamefont{Elhanine}},
  \bibinfo{author}{\bibfnamefont{B.}~\bibnamefont{Noller}},
  \bibinfo{author}{\bibfnamefont{P.}~\bibnamefont{Hemberger}},
  \bibinfo{author}{\bibfnamefont{I.}~\bibnamefont{Fischer}},
  \bibinfo{author}{\bibfnamefont{G.~A.} \bibnamefont{Garcia}},
  \bibinfo{author}{\bibfnamefont{H.}~\bibnamefont{Soldi-Lose}},
  \bibinfo{author}{\bibfnamefont{B.}~\bibnamefont{Gans}},
  \bibinfo{author}{\bibfnamefont{L.~A.} \bibnamefont{{Viera Mendes}}},
  \bibnamefont{et~al.}, \bibinfo{journal}{J.~Phys.~Chem.~A}
  \textbf{\bibinfo{volume}{114}}, \bibinfo{pages}{4818} (\bibinfo{year}{2010}).

\bibitem[{\citenamefont{Yu and Sears}(2002)}]{sears}
\bibinfo{author}{\bibfnamefont{H.-G.} \bibnamefont{Yu}} \bibnamefont{and}
  \bibinfo{author}{\bibfnamefont{T.}~\bibnamefont{Sears}},
  \bibinfo{journal}{J.~Chem.~Phys.} \textbf{\bibinfo{volume}{117}},
  \bibinfo{pages}{666} (\bibinfo{year}{2002}).

\bibitem[{\citenamefont{Keceli et~al.}(2009)\citenamefont{Keceli, Shiozaki,
  Yagi, and Hirata}}]{keceli}
\bibinfo{author}{\bibfnamefont{M.}~\bibnamefont{Keceli}},
  \bibinfo{author}{\bibfnamefont{T.}~\bibnamefont{Shiozaki}},
  \bibinfo{author}{\bibfnamefont{K.}~\bibnamefont{Yagi}}, \bibnamefont{and}
  \bibinfo{author}{\bibfnamefont{S.}~\bibnamefont{Hirata}},
  \bibinfo{journal}{Mol.~Phys.} \textbf{\bibinfo{volume}{107}},
  \bibinfo{pages}{1283} (\bibinfo{year}{2009}).

\bibitem[{\citenamefont{Ragni et~al.}(2016)\citenamefont{Ragni, Bitencourt,
  Prudente, Barreto, and Posati}}]{ch3harm}
\bibinfo{author}{\bibfnamefont{M.}~\bibnamefont{Ragni}},
  \bibinfo{author}{\bibfnamefont{A.~C.~P.} \bibnamefont{Bitencourt}},
  \bibinfo{author}{\bibfnamefont{F.~V.} \bibnamefont{Prudente}},
  \bibinfo{author}{\bibfnamefont{P.~R.~P.} \bibnamefont{Barreto}},
  \bibnamefont{and} \bibinfo{author}{\bibfnamefont{T.}~\bibnamefont{Posati}},
  \bibinfo{journal}{Eur. Phys. J. D} \textbf{\bibinfo{volume}{70}},
  \bibinfo{pages}{60} (\bibinfo{year}{2016}).

\bibitem[{\citenamefont{Reiher and Wolf}(2004{\natexlab{a}})}]{dkh1}
\bibinfo{author}{\bibfnamefont{M.}~\bibnamefont{Reiher}} \bibnamefont{and}
  \bibinfo{author}{\bibfnamefont{A.}~\bibnamefont{Wolf}},
  \bibinfo{journal}{J.~Chem.~Phys.} \textbf{\bibinfo{volume}{121}},
  \bibinfo{pages}{2037} (\bibinfo{year}{2004}{\natexlab{a}}).

\bibitem[{\citenamefont{Reiher and Wolf}(2004{\natexlab{b}})}]{dkh2}
\bibinfo{author}{\bibfnamefont{M.}~\bibnamefont{Reiher}} \bibnamefont{and}
  \bibinfo{author}{\bibfnamefont{A.}~\bibnamefont{Wolf}},
  \bibinfo{journal}{J.~Chem.~Phys.} \textbf{\bibinfo{volume}{121}},
  \bibinfo{pages}{10945} (\bibinfo{year}{2004}{\natexlab{b}}).

\bibitem[{\citenamefont{Hanauer and K{\"o}hn}(2009)}]{gecco1}
\bibinfo{author}{\bibfnamefont{M.}~\bibnamefont{Hanauer}} \bibnamefont{and}
  \bibinfo{author}{\bibfnamefont{A.}~\bibnamefont{K{\"o}hn}},
  \bibinfo{journal}{J.~Chem.~Phys.} \textbf{\bibinfo{volume}{131}},
  \bibinfo{pages}{124118} (\bibinfo{year}{2009}).

\bibitem[{\citenamefont{Hanauer and K{\"o}hn}(2011)}]{gecco2}
\bibinfo{author}{\bibfnamefont{M.}~\bibnamefont{Hanauer}} \bibnamefont{and}
  \bibinfo{author}{\bibfnamefont{A.}~\bibnamefont{K{\"o}hn}},
  \bibinfo{journal}{J.~Chem.~Phys.} \textbf{\bibinfo{volume}{134}},
  \bibinfo{pages}{204111} (\bibinfo{year}{2011}).

\bibitem[{\citenamefont{Ruden et~al.}(2004)\citenamefont{Ruden, Helgajer,
  J{\o}rgensen, and Olsen}}]{highorder1}
\bibinfo{author}{\bibfnamefont{T.~A.} \bibnamefont{Ruden}},
  \bibinfo{author}{\bibfnamefont{T.}~\bibnamefont{Helgajer}},
  \bibinfo{author}{\bibfnamefont{P.}~\bibnamefont{J{\o}rgensen}},
  \bibnamefont{and} \bibinfo{author}{\bibfnamefont{J.}~\bibnamefont{Olsen}},
  \bibinfo{journal}{J.~Chem.~Phys.} \textbf{\bibinfo{volume}{121}},
  \bibinfo{pages}{5874} (\bibinfo{year}{2004}).

\bibitem[{\citenamefont{Koput and Peterson}(2006)}]{highorder2}
\bibinfo{author}{\bibfnamefont{J.}~\bibnamefont{Koput}} \bibnamefont{and}
  \bibinfo{author}{\bibfnamefont{K.~A.} \bibnamefont{Peterson}},
  \bibinfo{journal}{J.~Chem.~Phys.} \textbf{\bibinfo{volume}{125}},
  \bibinfo{pages}{044306} (\bibinfo{year}{2006}).

\bibitem[{\citenamefont{Meier et~al.}(2011)\citenamefont{Meier, Neff, and
  Rauhut}}]{meier}
\bibinfo{author}{\bibfnamefont{P.}~\bibnamefont{Meier}},
  \bibinfo{author}{\bibfnamefont{M.}~\bibnamefont{Neff}}, \bibnamefont{and}
  \bibinfo{author}{\bibfnamefont{G.}~\bibnamefont{Rauhut}},
  \bibinfo{journal}{J.~Chem.~Theory Comput.} \textbf{\bibinfo{volume}{7}},
  \bibinfo{pages}{148} (\bibinfo{year}{2011}).

\bibitem[{\citenamefont{Rauhut et~al.}(2006)\citenamefont{Rauhut, Barone, and
  Schwerdtfeger}}]{chfclbr}
\bibinfo{author}{\bibfnamefont{G.}~\bibnamefont{Rauhut}},
  \bibinfo{author}{\bibfnamefont{V.}~\bibnamefont{Barone}}, \bibnamefont{and}
  \bibinfo{author}{\bibfnamefont{P.}~\bibnamefont{Schwerdtfeger}},
  \bibinfo{journal}{J.~Chem.~Phys.} \textbf{\bibinfo{volume}{125}},
  \bibinfo{pages}{054308} (\bibinfo{year}{2006}).

\bibitem[{\citenamefont{Rajam{\"a}ki et~al.}(2004)\citenamefont{Rajam{\"a}ki,
  Kallay, Noga, Valiron, and Halonen}}]{barrier_halonen}
\bibinfo{author}{\bibfnamefont{T.}~\bibnamefont{Rajam{\"a}ki}},
  \bibinfo{author}{\bibfnamefont{M.}~\bibnamefont{Kallay}},
  \bibinfo{author}{\bibfnamefont{J.}~\bibnamefont{Noga}},
  \bibinfo{author}{\bibfnamefont{P.}~\bibnamefont{Valiron}}, \bibnamefont{and}
  \bibinfo{author}{\bibfnamefont{L.}~\bibnamefont{Halonen}},
  \bibinfo{journal}{Mol. Phys.} \textbf{\bibinfo{volume}{102}},
  \bibinfo{pages}{2297} (\bibinfo{year}{2004}).

\bibitem[{\citenamefont{Rajam{\"a}ki et~al.}(2003)\citenamefont{Rajam{\"a}ki,
  Miani, and Halonen}}]{tunnel_halonen}
\bibinfo{author}{\bibfnamefont{T.}~\bibnamefont{Rajam{\"a}ki}},
  \bibinfo{author}{\bibfnamefont{A.}~\bibnamefont{Miani}}, \bibnamefont{and}
  \bibinfo{author}{\bibfnamefont{L.}~\bibnamefont{Halonen}},
  \bibinfo{journal}{J.~Chem.~Phys.} \textbf{\bibinfo{volume}{118}},
  \bibinfo{pages}{10929} (\bibinfo{year}{2003}).

\bibitem[{\citenamefont{Neff and Rauhut}(2014)}]{tunnel_neff}
\bibinfo{author}{\bibfnamefont{M.}~\bibnamefont{Neff}} \bibnamefont{and}
  \bibinfo{author}{\bibfnamefont{G.}~\bibnamefont{Rauhut}},
  \bibinfo{journal}{Spectrochim. Acta A} \textbf{\bibinfo{volume}{119}},
  \bibinfo{pages}{100} (\bibinfo{year}{2014}).

\bibitem[{\citenamefont{Richardson}(2017)}]{richardson}
\bibinfo{author}{\bibfnamefont{J.~O.} \bibnamefont{Richardson}},
  \bibinfo{journal}{Phys.~Chem.~Chem.~Phys.} \textbf{\bibinfo{volume}{19}},
  \bibinfo{pages}{966} (\bibinfo{year}{2017}).

\bibitem[{\citenamefont{Beyer et~al.}(2016)\citenamefont{Beyer, Richardson,
  Knowles, Rommel, and Althorpe}}]{beyer2016}
\bibinfo{author}{\bibfnamefont{A.~N.} \bibnamefont{Beyer}},
  \bibinfo{author}{\bibfnamefont{J.~O.} \bibnamefont{Richardson}},
  \bibinfo{author}{\bibfnamefont{P.~J.} \bibnamefont{Knowles}},
  \bibinfo{author}{\bibfnamefont{J.}~\bibnamefont{Rommel}}, \bibnamefont{and}
  \bibinfo{author}{\bibfnamefont{S.~C.} \bibnamefont{Althorpe}},
  \bibinfo{journal}{JCP} \textbf{\bibinfo{volume}{7}}, \bibinfo{pages}{4374}
  (\bibinfo{year}{2016}).

\bibitem[{\citenamefont{Richardson}(2018)}]{richardson2018}
\bibinfo{author}{\bibfnamefont{J.~O.} \bibnamefont{Richardson}},
  \bibinfo{journal}{J. Chem. Phys.} \textbf{\bibinfo{volume}{148}},
  \bibinfo{pages}{200901} (\bibinfo{year}{2018}).

\end{thebibliography}
\end{document}